\documentclass[fleqn,usenatbib]{mnras}
\usepackage{newtxtext,newtxmath}
\usepackage[T1]{fontenc}
\usepackage{ae,aecompl}
\usepackage{graphicx}
\usepackage{amsmath}
\usepackage{comment}
\usepackage{amssymb}
\usepackage{color}
\usepackage{multicol}  
\usepackage{bm} 
\usepackage{pdflscape}
\usepackage{cprotect} 
\usepackage{bigfoot} 
\usepackage[below]{placeins}

\def\apj{{\rm ApJ}}                 
\def\apjl{{\rm ApJ}}                
\def\apjs{{\rm ApJS}}               
\def\ao{{\rm Appl.~Opt.}}           
\def\aap{{\rm A\&A}}                
\def\mnras{{\rm MNRAS}}             
\def\prd{{\rm Phys.~Rev.~D}}        

\def\mnras{Mon.\ Not.\ R.\ Astron.\ Soc.\ }

\newcommand{\photoz}{photo-$z$}
\newcommand\salpha{{\eta_{\rm IA}}}
\newcommand{\bea}{\begin{eqnarray}}
\newcommand{\be}{\begin{equation}}
\newcommand{\ben}{\begin{enumerate}}
\newcommand{\bi}{\begin{itemize}}
\newcommand{\eea}{\end{eqnarray}}
\newcommand{\ee}{\end{equation}}
\newcommand{\ei}{\end{itemize}}
\newcommand{\een}{\end{enumerate}}

\newcommand{\matC}{\mathbf C}

\newcommand{\D}{\vek D}

\newcommand{\M}{\vek M}

\newcommand{\vek}{\mathbf}

\title[Interpreting Internal Consistency of DES Measurements]{Interpreting Internal Consistency of DES Measurements}
\author[V. Miranda, P. Rogozenski, and E. Krause,]{
V. Miranda,$^{1}$\thanks{E-mail: vivianmiranda@arizona.edu}
P. Rogozenski, $^{2}$
and E. Krause$^{1,2}$
\\
$^1$ Steward Observatory, Department of Astronomy, University of Arizona, Tucson, Arizona, 85721, USA
\\
$^2$ Department of Physics, University of Arizona, Tucson, Arizona, 85721, USA
}

\date{Accepted XXX. Received YYY; in original form ZZZ}
\pubyear{2020}

\begin{document}

\maketitle

\begin{abstract}
Bayesian evidence ratios are widely used to quantify the statistical consistency between different experiments. However, since the evidence ratio is prior dependent, the precise translation between its value and the degree of concordance/discordance requires additional information. The most commonly adopted metric, the Jeffreys scale, can falsely suggest agreement between datasets when priors are chosen to be sufficiently wide~\citep{Raveri:2018wln,Handley:2019wlz}. In this work, we examine evidence ratios in a DES-Y1 simulated analysis, focusing on the internal consistency between weak lensing and galaxy clustering. We study two scenarios using simulated data in controlled experiments. First, we calibrate the expected evidence ratio distribution given noise realizations around the best fit DES-Y1 $\Lambda$CDM cosmology. Second, we show the behavior of evidence ratios for noiseless fiducial data vectors simulated using a modified gravity model, which generates internal tension in the $\Lambda$CDM analysis. We show that the choice of prior could conceal the discrepancies between weak lensing and galaxy clustering induced by such models and that the evidence ratio in a DES-Y1 study is, indeed, biased towards agreement. 
\end{abstract}

\begin{keywords}
cosmological parameters -- theory -- large-scale structure of the Universe
\end{keywords}

\renewcommand{\thefootnote}{\arabic{footnote}}
\setcounter{footnote}{0}

\section{Introduction}
\quad \quad Since the discovery of the accelerating expansion of the universe~\citep{Riess:1998cb,Perlmutter:1998np}, various surveys have been designed to measure the background expansion and structure formation of the Universe with increasing precision. The Dark Energy Task Force (DETF)~\citep{Albrecht:2006um} classifies these surveys from stage I to stage IV according to their ability to increase the figure-of-merit~\citep{2009arXiv0901.0721A} of the $w_0-w_a$ parameterization for the dark energy equation of state~\citep{Linder:2002et,Chevallier:2000qy}. The community is currently analyzing the stage III surveys, while stage IV surveys such as DESI~\citep{Levi:2019ggs}, Nancy Grace Roman Space Telescope~\citep{2019arXiv190205569A}, CMB-S4~\citep{Abazajian:2016yjj} and Vera Rubin Telescope Legacy Survey of Space and Time (LSST)~\citep{DESC_SRD18} will start collecting data in the next few years with the potential to significantly expand our knowledge about the early and late-time cosmos.  

Ongoing stage III surveys, such as the Dark Energy Survey (DES)~\citep{Abbott:2005bi}, constrain the parameters of the standard model ($\Lambda$CDM) with unprecedented precision. These constraints encompass measurements of the Cosmic Microwave Background (CMB)~\citep{2018arXiv180706205P,2012SPIE.8452E..1EA,2016ApJS..227...21T}, Type Ia Supernova~\citep{2019ApJ...872L..30A,Scolnic:2017caz}, Baryon Acoustic Oscillations (BAO)~\citep{Alam:2016hwk,Prakash:2015trd,DAmico:2019fhj,Ivanov:2019hqk}, Weak Lensing~\citep{Asgari:2020wuj,Hildebrandt:2018yau,Troxel:2017xyo,Hikage:2018qbn} and Galaxy Clustering~\citep{2015MNRAS.452.2087L,Elvin-Poole:2017xsf,DAmico:2019fhj,Ivanov:2019hqk}. As demonstrated by the DETF, combining probes is the most promising way forward to make more meaningful statements on the physical properties of dark energy~\citep{Albrecht:2006um,Abbott:2017wau,2020arXiv200715632H,Tegmark:1998yy,Eisenstein:1998tu,Tegmark:1998ab}.

However, constraining cosmological parameters through combining multiple probes requires that these are statistically consistent. The presence of tensions with moderate to high statistical significance, which can prevent datasets from being integrated, have been observed in stage III experiments~\citep{Riess:2019cxk,Douspis:2018xlj,DiValentino:2019qzk,Abbott:2017smn,Hikage:2018qbn,2020arXiv200715632H}. In particular, the current inconsistency between CMB predictions~\citep{Aghanim:2018eyx} and local measurements of the Hubble constant, $H_0$,~\citep{Riess:2019cxk} is a good example of a tension that may require new physics to be fully resolved ~\citep{Knox:2019rjx,Verde:2019ivm}. 

The Dark Energy Survey uses the combination of weak lensing and galaxy clustering to break degeneracies between dark energy and other parameters. For example, the DES year one (DES-Y1) error bars from the cosmic shear investigation on the dark energy equation of state are reduced by $\sim 30\%$ in the combined analysis~\citep{Abbott:2017wau, Troxel:2017xyo}. The joint analysis is only permitted however if the datasets are statistically consistent. In \cite{Abbott:2017wau}, consistency was ascertained by the Bayesian evidence ratio, $R$, utilizing the Jeffreys Scale. However, analytical examples show that the Jeffreys scale should not be used as an universal scale~\citep{Nesseris:2012cq}, given that priors can always be chosen to be wide enough to enable consistency~\citep{Marshall:2004zd}.

In order to make meaningful statements about consistency of datasets it is important to investigate how the Bayesian evidence ratio $R$ is affected by the priors under consideration. These investigations are particularly relevant when tension with modest statistical significance is detected, e.g., the disagreement between Planck data and weak lensing surveys over the value of $S_8 \equiv  \sigma_8 \Omega_m^{1/2}$ parameter~\citep{Abbott:2017wau,2020arXiv200715632H,Hikage:2018qbn}.

Given the demanding computational costs associated with Bayesian evidence computation~\citep{2015MNRAS.453.4384H}, calibrating survey data concordance with simulated data is not always feasible. Alternative metrics with reduced prior dependence have been suggested~\citep{Handley:2019wlz,Seehars:2015qza}. In simple cases (e.g. multivariate Gaussians), these alternatives can be prior independent. However, in more general cases, the interpretation of alternative metrics still requires careful scale calibration using simulated data. Yet another approach to reduce prior dependencies is to adopt approximations, such as the validity of the Gaussian linear model (GLM), which allows Bayesian estimators to be computed either analytically or from Monte Carlo Markov Chains~\citep{Raveri:2018wln}.

In this paper we examine the Bayesian evidence ratio in the context of quantifying consistency between cosmic shear, galaxy-galaxy lensing, and galaxy clustering in DES-Y1 data. In particular, we want to quantify whether cosmic shear and the combination of galaxy clustering and galaxy-galaxy lensing (so-called 2x2pt) can be combined into a so-called 3x2pt analysis. We test how this metric responds to noise drawn from the DES-Y1 covariance around the best-fit cosmology at varying confidence intervals in $\vec \chi^2$ space. This first test demonstrates how 
`real' survey noise at known deviations from the best-fit cosmology propagates into Bayesian estimators. We then explore how the evidence ratio behaves when data vectors generated from an underlying modified gravity theory are fit with the standard model. When confined to the standard model, these modified gravity based data vectors naturally induce a tension between weak lensing and galaxy clustering. 

This manuscript is structured as follows: In Sect.~\ref{sec:tension_metric} we define the tension metrics studied in this paper. In Sect.~\ref{sec::setup} we explain the theoretical modeling and aspects of our simulated analyses. Section \ref{sec:lcdm} describes our findings about Bayesian evidence ratios and other tension metrics when considering noisy $\Lambda$CDM data vectors that are analyzed with a $\Lambda$CDM model. This scenario corresponds to the case where realistic noise in a data vector might be misinterpreted as a physical tension. In Sect. 5 we consider a noise free modified gravity data vector that is analyzed with a $\Lambda$CDM model. This scenario mimics the case where an actual physical tension between the clustering and weak lensing parts of the data vector exist. Four appendices offer further explanation of the details that are only summarized in this section. We conclude in Sect.~\ref{sec:conclusion}. 

\section{Tension Metric Definitions}
\label{sec:tension_metric}

\quad\quad In this section we briefly review tension metrics and establish consistent notation. We start defining the posterior probability for a set of parameters $\vec{\theta}$ in a given model $\mathcal{H}$ and observed dataset $d$ as $P(\vec{\theta} | d, \mathcal{H})$. The posterior is related to the likelihood, $P(d | \vec{\theta}, \mathcal{H})$, via the Bayes' Theorem
\begin{align}
    P(\vec{\theta} | d, \mathcal{H}) = \frac{P(d | \vec{\theta}, \mathcal{H}) P(\vec{\theta} | \mathcal{H})}{P(d | \mathcal{H})} \,.
\end{align}
The prior, $P(\vec{\theta} | \mathcal{H})$, describes the \textit{a priori} probability distribution of the parameters $\vec{\theta}$ within the assumed  model $\mathcal{H}$. The normalization factor, $P(d | \mathcal{H})$, is called the Bayesian evidence~\citep{Marshall:2004zd}. 

\subsection{Bayesian Evidence Ratio}\label{sec:evidence ratio}
\quad\quad The Bayesian evidence of M datasets $\vec d = (d_1,\dotso,d_M)$ given a model $\mathcal{H}$ of N parameters $\vec \theta = (\theta_1,\dotso,\theta_N)$ is given by
\begin{align}
P(\vec d | \mathcal{H}) = \int d\vec \theta P(\vec d | \vec \theta, \mathcal{H}) P(\vec \theta | \mathcal{H}) \, . 
\end{align}
In order to evaluate the probability that experiments $d_1$ and $d_2$ are in agreement, we evaluate the odds of hypothesis $\mathcal{H}_0$, that we can model both datasets with a single set of parameters, against the alternative hypothesis $\mathcal{H}_1$, that modeling each dataset with a different set of parameters is preferable. 

These odds are defined as $\mathcal{P}(\mathcal{H}_0 | d_1, d_2)/\mathcal{P}(\mathcal{H}_1 | d_1, d_2)$ and their relation to the evidences $P(d_1, d_2 | \mathcal{H}_0)$ and $P(d_1, d_2 | \mathcal{H}_1)$ can be readily seen when applying Bayes' theorem
\begin{align}\label{eqn:ratio_model_posterior}
\frac{\mathcal{P}(\mathcal{H}_0 | d_1, d_2)}{\mathcal{P}(\mathcal{H}_1 | d_1, d_2)} = \frac{P(d_1, d_2 | \mathcal{H}_0)}{P(d_1, d_2 | \mathcal{H}_1)}  \cdot \frac{P(\mathcal{H}_0)}{P(\mathcal{H}_1)} \,,
\end{align}
where $P(H_{i=\{0,1\}})$ are the prior probabilities of models $\mathcal{H}_{i=\{0,1\}}$. The first ratio on the right-hand side of Eq.~\ref{eqn:ratio_model_posterior} is known as the Bayesian evidence ratio, R. If the datasets are independent, we may express it as 
\begin{align}
\label{eqn:evidence_ratio}
R =  \frac{P(d_1, d_2 | \mathcal{H}_0)}{P(d_1 | \mathcal{H}_1) P(d_2 | \mathcal{H}_1) } \, .
\end{align}

The Bayesian evidence ratio generally implies agreement between datasets when $R \gg 1$, while $R \ll 1$ flags the opposite. The ratio changes as a function of prior range, which can mimic consistency even in the presence of tension.

\subsection{$\Delta\bar{\chi}^2$ statistic}
\label{subsec::delta-chi}
\quad \quad The $\bar{\chi}^2$ value is a statistic related to the average log-likelihood of a chain marginalized over the posterior. Given the weights of each sample $i$ of a chain of length $N$, we calculate the statistic directly as  
\begin{align}
   \bar{\chi}_j^2 = -2\big\langle \ln P(\vec d_j | \vec \theta, \mathcal{H}) \big\rangle = -2\frac{\sum_i^N w_i \ln P_i(\vec d_j | \vec \theta, \mathcal{H})}{\sum_i^N w_i},
\end{align}
where the sample weights are defined as the ratio of the sample posterior over the maximum sampled posterior of the chain. We define a statistic similar to the delta chi-squared statistic of~\citep{Marshall:2004zd} as the difference between the $\bar{\chi}^2$ values of the joint and independent datasets as:
\begin{align}\label{subsec::delta-chi-eq}
    \Delta\bar{\chi}^2 &= \bar{\chi}_{12}^2 - (\bar{\chi}_{1}^2 + \bar{\chi}_{2}^2) \,.
\end{align}

\subsection{Generalized Parameter Distance}

\quad \quad The Generalized Parameter Distance estimates the departure from the fiducial vector (in this case determined by the DES-Y1 best-fit cosmology) and it is determined by calculating the covariance of a chain, $\hat{\Sigma}$, then taking the difference, in parameter space,  of the fiducial data vector, $\vec{\mu}$ and the best-fit data vector of the samples, $\vec{\theta}$, as 
\begin{align}
   \Delta \equiv \sqrt{\big(\vec{\theta} - \vec{\mu}\big)^t\hat{\Sigma}^{-1}\big(\vec{\theta} - \vec{\mu}\big)},
\end{align}

\subsection{The Kullback-Leibler Divergence}

\quad \quad Alternatively to the evidence ratio, the Kullback-Leibler (KL) Divergence, also known as the relative entropy, determines how parameters are constrained by the data compared to the prior constraints~\citep{kullback1951}. Defined as
\begin{align}
\mathcal{D}_i = \int d\vec \theta P(\vec \theta | d_i, \mathcal{H}) \ln \Bigg[ \frac{P(\vec \theta | d_i, \mathcal{H})}{P(\vec \theta | \mathcal{H})}\Bigg] \, , 
\end{align}
the  KL Divergence is invariant under model reparameterization and can be interpreted as measuring the information gain when going from the prior distribution to the posterior. Similar to entropy, $D_i \geq 0$. The KL Divergence can also measure the information gain of augmented datasets by taking $P(\vec \theta | \mathcal{H}) \rightarrow P(\vec \theta | d_i, \mathcal{H})$ and $P(\vec \theta | d_i, \mathcal{H}) \rightarrow P(\vec \theta | d_i + d_{new}, \mathcal{H})$. The relative entropy between datasets is the basis of a tension metric called Surprise~\citep{Seehars:2014ora,Seehars:2015qza}. Both the KL Divergence and Surprise computation is non-trivial outside the Gaussian case, which limits their applicability as a check for statistical consistency. 

\subsection{Suspiciousness}

\quad \quad Suspiciousness is a tension metric that aims to alleviate the prior dependence exhibited in the evidence ratio~\citep{Handley:2019wlz}. This metric is defined as   
\begin{align}\label{eq: sus}
\ln S \equiv \ln R - \ln I\,,
\end{align}
where $\ln I$ is defined as the information ratio
\begin{align}
\ln I \equiv \mathcal{D}_1 + \mathcal{D}_2 - \mathcal{D}_{12}. 
\end{align}
In restricted cases (e.g. the case of flat priors imposed on a multivariate Gaussian likelihood), the prior dependence in the metric is completely eliminated. For this particular case, a generalization to correlated datasets has been found~\cite{Lemos:2019txn}. Details on the numerical evaluation of suspiciousness, as well as the evidence, in a nested sampling run are shown in Appendix~\ref{section:apxD}.

\section{Modeling and Analysis Choices}
\label{sec::setup}
\quad \quad The theoretical modeling and covariance computation and validation for the DES-Y1 3x2pt analysis are described in detail in \citep{Y1Methods}. We summarize the main modeling details briefly below.

\subsection{Modeling Details - Observables}
\label{sec:model_obs}

The DES 3x2pt data vector consists of the angular galaxy clustering statistic $w^i(\theta)$ of galaxies in redshift bin $i$, the galaxy--galaxy lensing statistic $\gamma_\mathrm{t}^{ij}(\theta)$ for galaxies in redshift bin $i$ and shape measurements for source galaxies in redshift bin $j$, and cosmic shear two-point correlations functions $\xi_{\pm}^{ij}(\theta)$ of shape measurements for source galaxies in redshift bins $i,j$. The galaxy sample used in the clustering measurement, which also constitutes the ``lens'' sample for galaxy-galaxy lensing, is selected using the redMaGiC algorithm \citep{rra15}. Details on the DES-Y1 sample selection and redshift calibration described in \citet{Elvin-Poole:2017xsf,2018MNRAS.481.2427C}. For the weak lensing galaxy sample, we adopt the DES-Y1 \textsc{metacal} source galaxy sample, for which the sample selection from the DES-Y1 gold catalog \citep{y1gold} and the shear catalog are described in \citet{y1shearcat}, and the source redshift estimates are described in \citet{y1photoz}, respectively.

\begin{table}
\caption{Table with priors for the cosmological and nuisance parameters, similar to the adopted priors in DES-Y1. In addition, we
applied flat($0.005, 0.04$) priors on $\Omega_b h^2$ for minimal compatibility with BBN constraints in CosmoLike (see Appendix~\ref{sec::appendixA} for further details). }
\begin{center}
\begin{tabular*}{0.4\textwidth}{@{\extracolsep{\fill}}| c c |}
\hline
\hline
Parameter & Prior \\  
\hline 
\multicolumn{2}{|c|}{{\bf Cosmology}} \\
$\Omega_m$  &  flat ($0.10, 0.90$)  \\ 
$A_s \times 10^{-9}$ &  flat ($0.50,5.00$)  \\ 
$n_s$ &  flat (0.87, 1.07)  \\
$\Omega_b$ &  flat (0.03, 0.07)  \\
$H_0$  &  flat (55.0, 91.0)   \\
$m_\nu$  & flat($0.06$, $0.93$) \\
\multicolumn{2}{|c|}{{\bf Lens Galaxy Bias}} \\
$b_{i} (i=1,5)$   & flat (0.8, 3.0) \\
\hline
\multicolumn{2}{|c|}{{\bf Intrinsic Alignment}} \\
\multicolumn{2}{|c|}{{$A_{\rm IA}(z) = A_{\rm IA} [(1+z)/1.62]^\salpha$}} \\
$A_{\rm IA}$   & flat ($-5,5$) \\
$\salpha$   & flat ($-5,5$) \\
\hline
\multicolumn{2}{|c|}{{\bf Lens \photoz\ shift}} \\
$\Delta z^1_{\rm l}$  & Gauss ($0, 0.008$) \\
$\Delta z^2_{\rm l}$  & Gauss ($0, 0.007$) \\
$\Delta z^3_{\rm l}$  & Gauss ($0, 0.007$) \\
$\Delta z^4_{\rm l}$  & Gauss ($0, 0.010$) \\
$\Delta z^5_{\rm l}$  & Gauss ($0, 0.010$) \\
\hline
\multicolumn{2}{|c|}{{\bf Source \photoz\ shift}} \\
$\Delta z^1_{\rm s}$  & Gauss ($0, 0.015$) \\
$\Delta z^2_{\rm s}$  & Gauss ($0, 0.013$) \\
$\Delta z^3_{\rm s}$  & Gauss ($0, 0.011$) \\
$\Delta z^4_{\rm s}$  & Gauss ($0, 0.022$) \\
\hline
\multicolumn{2}{|c|}{{\bf Shear calibration}} \\
$m^{i}(i=1,4)$ & Gauss ($0, 0.023$)\\
\hline
\end{tabular*}
\end{center}
\label{tab:params}
\end{table}

We denote the redshift distribution of the redMaGiC/Metacal source galaxy sample in tomography bin $i$ as $n_{\mathrm{g}/\kappa}^i(z)$, and the angular number densities of galaxies in this redshift bin as 
\begin{equation}
\bar{n}_{\mathrm{g}/\kappa}^i = \int dz\; n_{\mathrm{g}/\kappa}^i(z)\,.
\end{equation}
Assuming a flat $\Lambda$CDM universe, we write the radial weight function for clustering in terms of the comoving radial distance $\chi$ as
\begin{equation}
q_{\delta_{\mathrm{g}}}^i(k,\chi) = b^i\left(k,z(\chi)\right)\frac{n_{\mathrm{g}}^i(z(\chi)) }{\bar{n}_{\mathrm{g}}^i}\frac{dz}{d\chi}\,,
\end{equation}
with $b^i(k,z(\chi))$ the galaxy bias of the redMaGiC galaxies in tomography bin $i$,
and the lensing efficiency 
\begin{equation}
q_\kappa^{i}(\chi) = \frac{3 H_0^2 \Omega_m }{2 \mathrm{c}^2}\frac{\chi}{a(\chi)}\int \mathrm d \chi' \frac{n_{\kappa}^{i} (z(\chi')) dz/d\chi'}{\bar{n}_{\kappa}^{i}} \frac{\chi'-\chi}{\chi'} \,,
\end{equation}
where $H_0$ is the Hubble constant, $c$ the speed of light, and $a$ the scale factor.
The angular power spectra for cosmic shear, galaxy-galaxy lensing, and galaxy clustering are calculated using the Limber approximation
\begin{align}
\label{eq:Cell}
\nonumber C_{\kappa \kappa}^{ij} (l) &=\!\! \int\!\! d\chi \frac{q_\kappa^i(\chi) q_\kappa^j(\chi) }{\chi^2} P_{\mathrm{NL}}\!\!\left(\frac{l+1/2}{\chi},z(\chi)\right),\\
\nonumber C_{\delta_{\mathrm{g}}\kappa}^{ij}(l) &=\!\! \int\!\! d\chi\! \frac{q_{\delta_{\mathrm{g}}}^i\!\!\left(\frac{l+1/2}{\chi},\chi\right) q_\kappa^j(\chi)}{\chi^2}  P_{\mathrm{NL}}\!\left(\frac{l+1/2}{\chi},z(\chi)\right),\\
C_{\delta_{\mathrm{g}}\delta_{\mathrm{g}}}^{ij}(l) &=\!\! \int\!\! d\chi\! \frac{q_{\delta_{\mathrm{g}}}^i\!\!\left(\!\frac{l+1/2}{\chi},\chi\right)q_{\delta_{\mathrm{g}}}^j\!\left(\frac{l+1/2}{\chi},\chi\right)}{\chi^2}  P_{\mathrm{NL}}\!\!\left(\frac{l+1/2}{\chi},z(\chi)\!\!\right)\, ,
\end{align}
where $P_{\mathrm{NL}}(k,z)$ is the non-linear matter power spectrum at wave vector $k$ and redshift $z$ computed via  \verb'Halofit' \citep{Takahashi:2012em}.

The angular correlation functions are calculated from the angular power spectra as
\begin{align}
\nonumber \xi_{+/-}^{ij}(\theta) &= \int \frac{dl\, l}{2\pi} J_{0/4}(l\theta) C_{\kappa \kappa}^{ij}(l)\,,\\
\nonumber \gamma_{\mathrm{t}}^{ij}(\theta) &=\int \frac{dl\, l}{2\pi} J_2(l\theta) C_{\delta_{\mathrm{g}}\kappa}^{ij}(l) \,,\\
w^{i}(\theta)& = \sum_l \frac{2l+1}{4\pi} P_l\left(\cos(\theta)\right)\,C_{\delta_{\mathrm{g}}\delta_{\mathrm{g}}}^{ii}(l)\,,
\end{align}
with $J_n(x)$ the $n$-th order Bessel function of the first kind, and $P_l(x)$ the Legendre polynomial of order $l$.

\subsection{Modeling Details - Systematics}
\label{sec:model_sys}
The DES-Y1 baseline model includes nuisance parameters to account for uncertainties in astrophysical and observational systematic effects, summarized below. Prior distributions of our parameters are given in Table \ref{tab:params}, similar to those in DES-Y1 analyses. Parameters with Gaussian priors (i.e. the lens photo-$z$ shifts, the source photo-$z$ shifts, and the shear calibrations) are prior-dominated. A detailed validation of these parameterizations can be found in \cite{Elvin-Poole:2017xsf, Y1Methods} and \cite{Troxel:2017xyo}.
\paragraph*{Photometric redshift uncertainties} 
The uncertainty in the redshift distribution $n$ is modeled through shift parameters $\Delta_z$,
\begin{align}
n^i_{x}(z) = \hat{n}^i_{x}\left(z-\Delta^i_{z,x}\right)\,,\;\;\; x\in\left\{\mathrm{g},\kappa\right\}\,,
\end{align}
where $\hat{n}$ denotes the estimated redshift distribution. We marginalize over one parameter for each source and lens redshift bin (nine parameters in total), using the the priors derived in \cite{y1photoz,2018MNRAS.481.2427C}.
\paragraph*{Multiplicative shear calibration} is marginalized using one parameter $m^i$ per redshift bin, which affects cosmic shear and galaxy--galaxy lensing correlation functions via
\begin{align}
\nonumber \xi_\pm^{ij}(\theta) \quad &\longrightarrow& \quad (1+m^i) \, (1+m^j) \, \xi_\pm^{ij}(\theta), \\
\gamma_t^{ij}(\theta) \quad &\longrightarrow& \quad (1+m^j) \, \gamma_t^{ij}(\theta),
\label{eq:m}
\end{align}
with Gaussian priors as determined in \cite{Troxel:2017xyo,y1shearcat}.
\paragraph*{Galaxy bias} The DES-Y1 baseline model assumes an effective linear galaxy bias ($b_{{1}}$) using one parameter per galaxy redshift bin $b^i(k,z) = b_1^i$,
i.e. five parameters, which are marginalized over conservative flat priors.
\paragraph*{Intrinsic galaxy alignments} (IA) are modeled using a power spectrum shape and amplitude $A(z)$, assuming the non-linear linear alignment (NLA) model \citep{his04,brk07} for the IA power spectrum. The impact of this specific IA power spectrum model can be written as  
\begin{align}
\label{eq:CIA}
q_\kappa^{i}(\chi) \quad &\longrightarrow&q_\kappa^{i}(\chi) - A\left(z\left(\chi\right)\right)\frac{n^i_{\kappa}(z(\chi))}{\bar{n}^i_{\kappa}} \frac{dz}{d \chi}\,.
\end{align}
The IA amplitude is modeled as a power-law scaling in $(1+z)$ with normalization $A_{\mathrm{IA,0}}$ and power law slope $\alpha_{\mathrm{IA}}$, which are both marginalized using conservative priors.

\begin{figure}
\includegraphics[width=0.45\textwidth,height=0.42\textwidth]{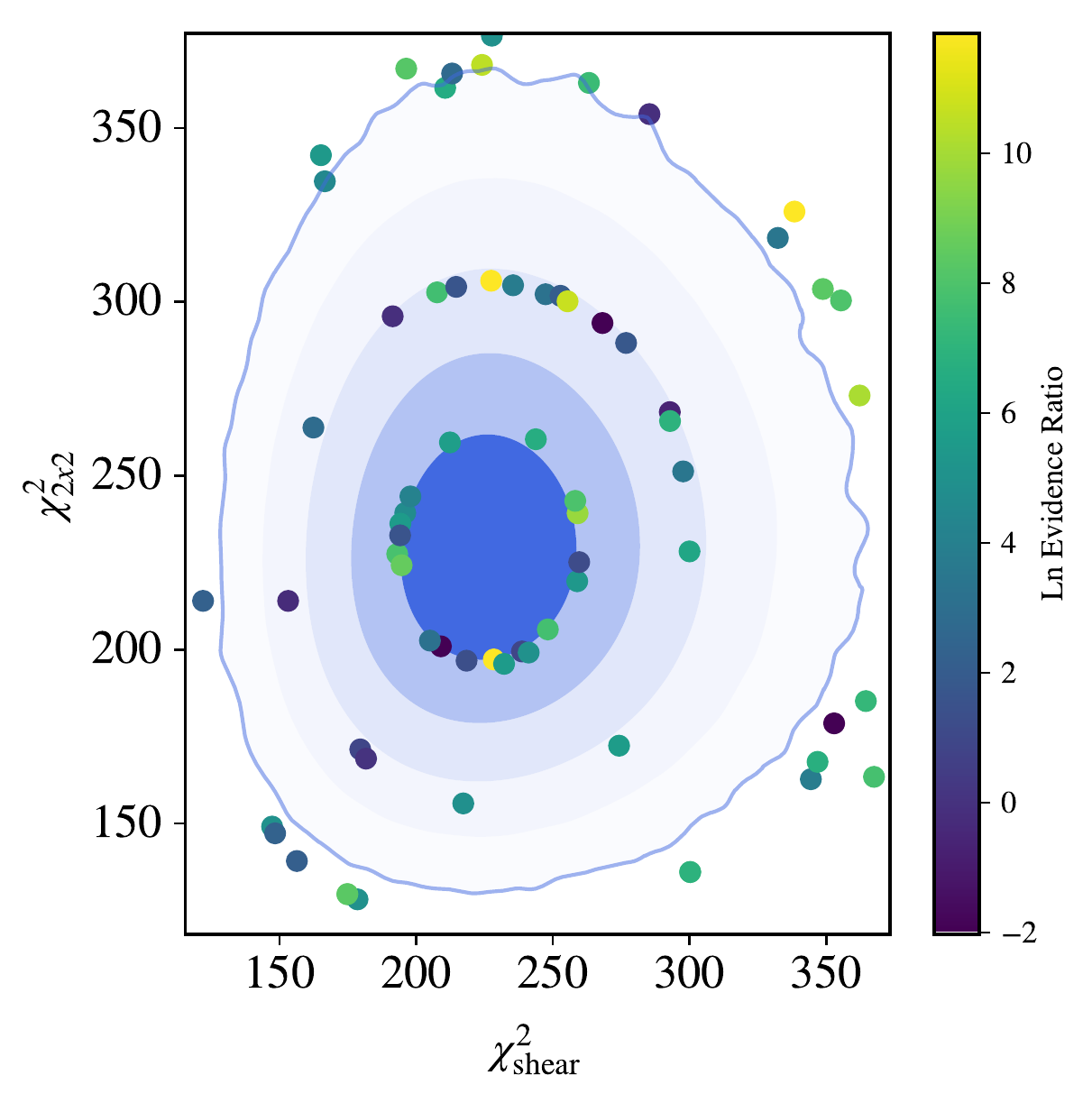}
\cprotect\caption{The distribution of $\vec \chi^2$ for cosmic shear, $\chi^2_\text{shear}$, and the 2x2pt (galaxy-galaxy lensing and galaxy clustering), $\chi^2_\text{2x2}$, generated using the DES-Y1 joint covariance matrix. We compute the $68\%$, $99.7\%$, and $99.99997\%$ confidence intervals from the generation of hundreds of millions of noise realizations,  smooth the contours, and define confidence intervals using a KDE. The data vectors are chosen along these contours and are represented as colored points. The color-code denotes the log-evidence ratio of the 3x2pt evidence to the 2x2pt and shear evidences (c.f. Eq. \ref{eqn:evidence_ratio}). Our selected points are sample the confidence limits in all radial directions and we don't find radial or angular trends of the evidence ratio.}
\label{fig:evidence_ratio1}
\end{figure}

\section{Evidence Ratio as a function of noisy $\Lambda$CDM data vectors}
\label{sec:lcdm}
\quad\quad In this section, we analyze the distribution of Bayesian evidence ratios for a set of realistic noise realizations of the DES-Y1 data vectors around the DES-Y1 best-fit $\Lambda$CDM cosmology. We aim to examine which of these noise realizations of $\Lambda$CDM can be flagged as tension according to the Jeffreys scale. We also investigate whether noise realizations at the one $\sigma$ level are more or less likely to be classified as tension by the Jeffreys scale compared to three and five sigma events. 

\begin{figure}
\includegraphics[width=0.45\textwidth,height=0.38\textwidth]{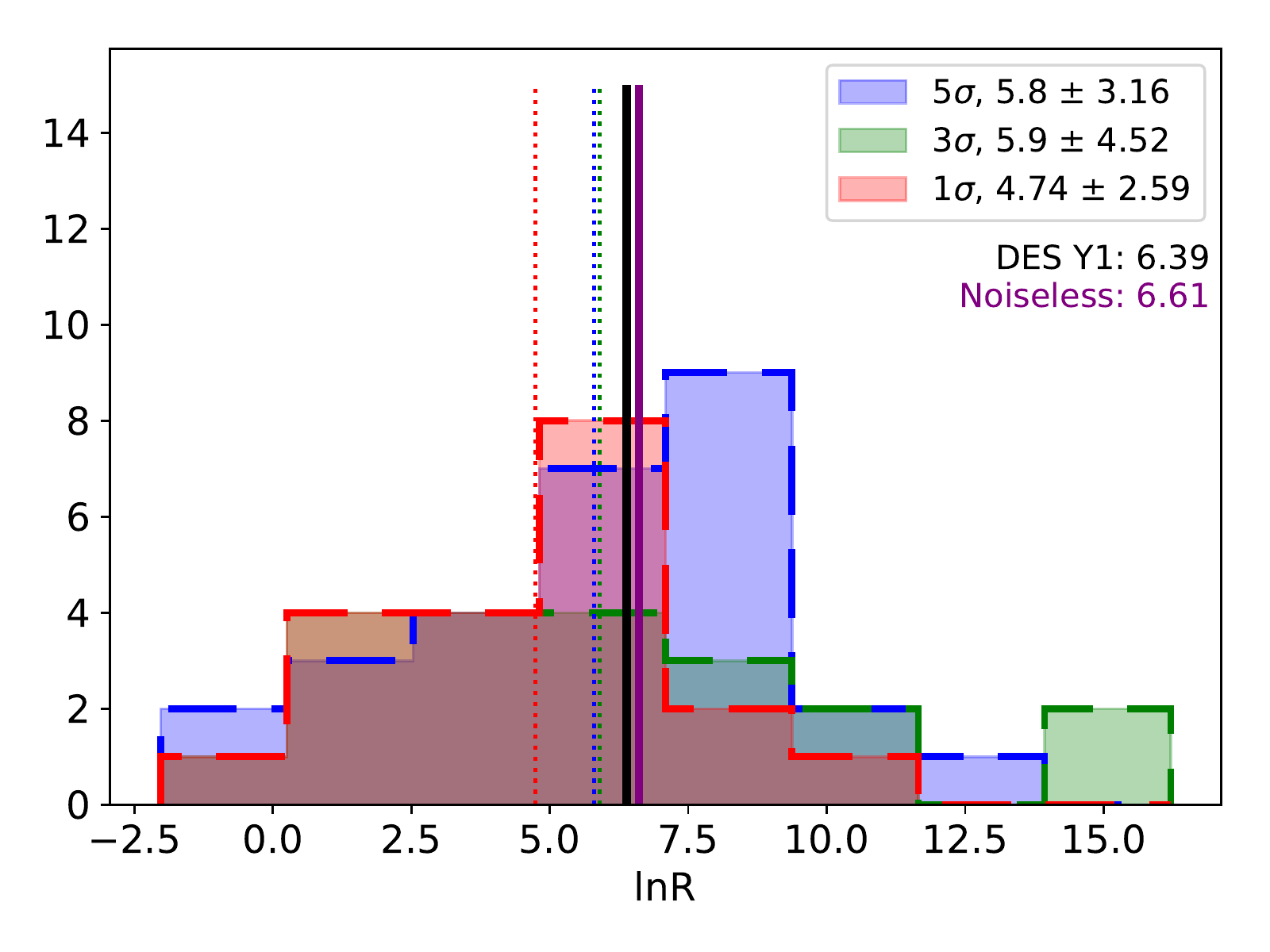}
\cprotect\caption{Histogram of evaluated log-evidence at the one, three, and five $\sigma$ confidence intervals. For comparison, we include the log-evidence ratios of our noiseless fiducial data vector and the official DES-Y1 analysis. The mean log-evidence ratio of each confidence interval is represented as a dotted line, with the mean and scatter explicitly given for each interval in the top-right key. The histogram reveals that the points on each contour all have similar log-evidence ratio distributions. The histogram also shows that the observed DES-Y1 evidence ratio is rather typical and does not point to an unusual level of agreement between the datasets, where the Jeffreys scale declares the DES-Y1 log-evidence ratio to be decisive agreement.}
\label{fig:evidence_ratio_hist}
\end{figure} 

\begin{figure*}
\includegraphics[width=0.46\textwidth,height=0.35\textwidth]{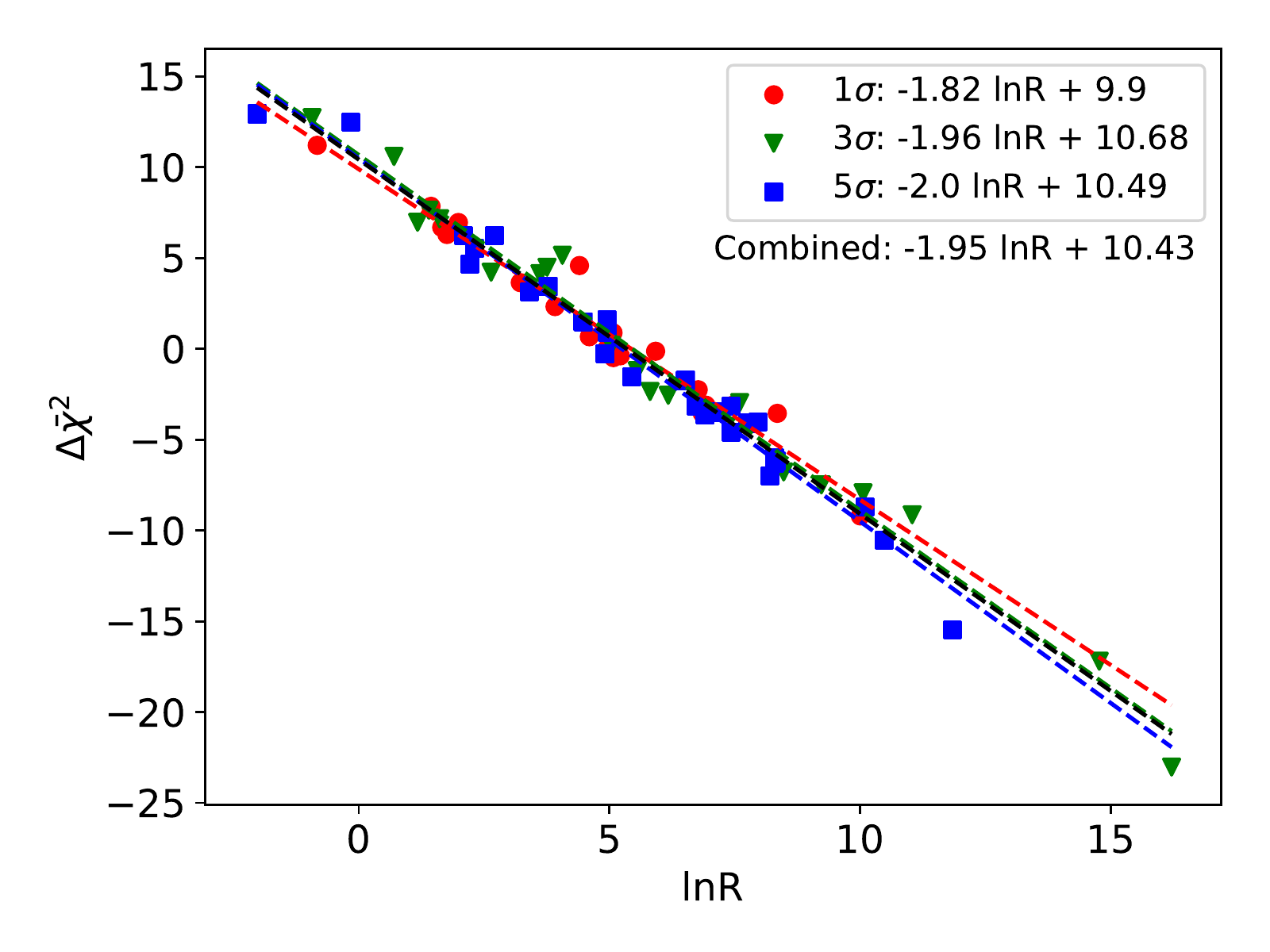}
\includegraphics[width=0.46\textwidth,height=0.35\textwidth]{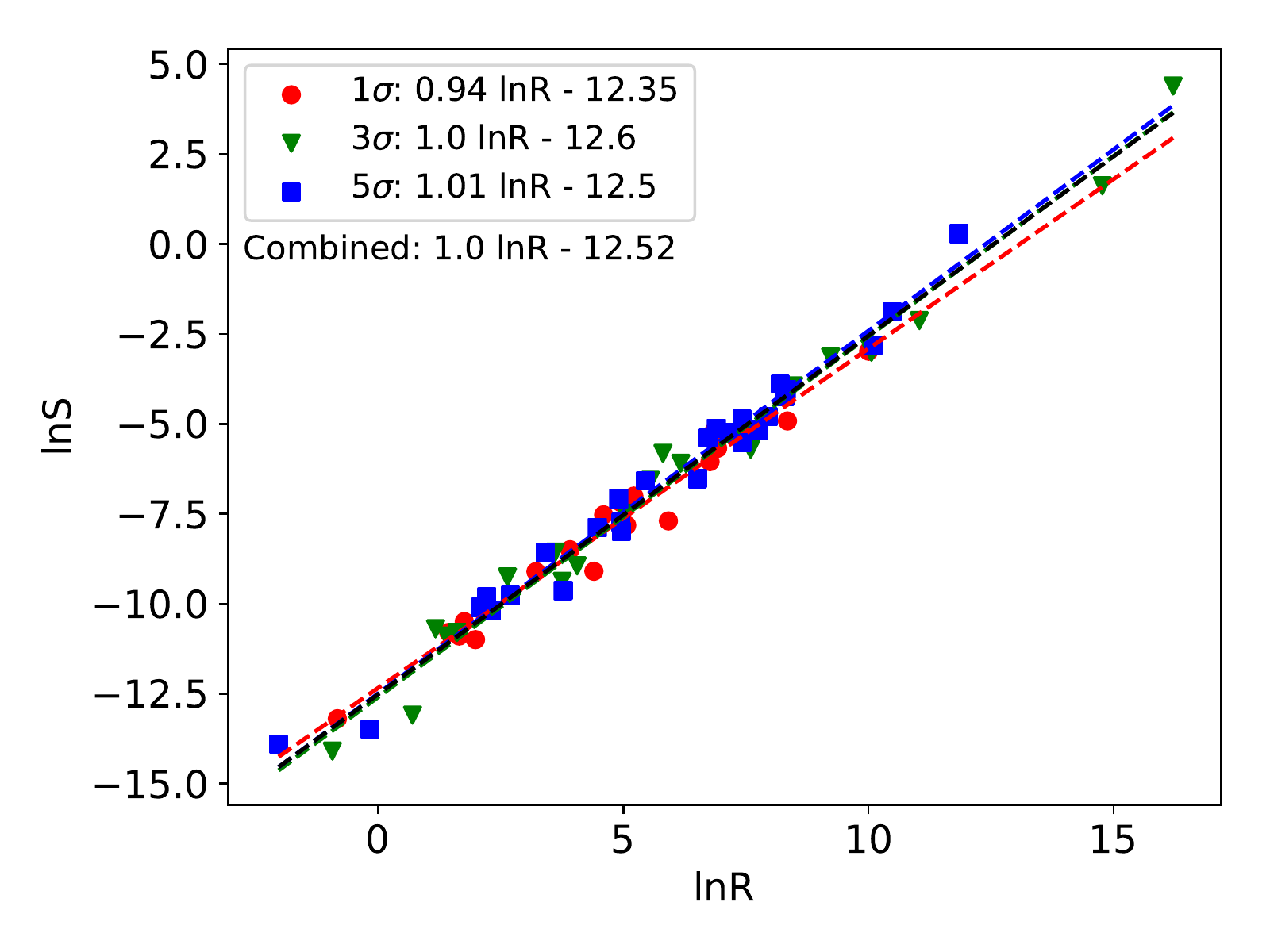}
\cprotect\caption{Correlation between Bayesian evidence ratios and $\Delta\bar{\chi}^2$ (left panel), Bayesian evidence ratios and suspiciousness (right panel). In both cases, the fit parameters of the slope are similar for one, three, and five $\sigma$ noise realizations. For $\Delta\bar{\chi}^2$, the slope of the fit is close to the predicted for multivariate Gaussian posteriors.}
\label{fig:evidence_ratio2}
\end{figure*} 

\subsection{Likelihood Analysis}
\label{sec:like}
In the following two sections we run multiple simulated DES-Y1 likelihood analyses to explore the distribution of Bayesian evidence ratios as a function different input data vectors. The input data vectors computed in Sect. \ref{sec:lcdm_data} resemble realistic noise realizations of the DES-Y1 survey assuming the DES-Y1 best-fit cosmology. The input data vectors in Sect. \ref{sec:mg_data} are computed from a modified gravity model, thereby inducing a physical tension between the weak lensing and the galaxy clustering part of the data vector. 

Throughout this paper we assume that the likelihood function ($\mathcal{L}$) of our data vector ($\D$) is well approximated by a multivariate Gaussian
\begin{align}
\label{eq:like}
\mathcal{L} \propto \exp \biggl(-\frac{1}{2} \left[ \left(\D -\M(\vec \theta)\right)^t \, \matC^{-1} \, \left(\D-\M(\vec \theta)\right) \right]  \biggr) \, ,
\end{align}
where $\M$ denotes the theory prediction or model vector.
As ~\cite{2019arXiv190503779L} demonstrate Gaussian functional form is a acceptable approximation, at least for ongoing and future cosmic shear surveys.

We use \verb'CosmoLike' \citep{kre17} with \verb+CLASS+ \citep{2011arXiv1104.2932L,2011JCAP...07..034B,2011arXiv1104.2934L,2011JCAP...09..032L} to compute the fiducial data vector and covariance. We sample the parameter space with the \verb'Polychord' \citep{2015MNRAS.453.4384H} nested sampling, with an interface implemented in the \verb'Cobaya' framework \citep{Torrado:2020dgo}, assuming the \verb+CAMB+~\citep{Lewis:1999bs,Howlett:2012mh} Boltzmann code. We perform extensive tests of our pipeline that merged \verb'CosmoLike' and \verb'Cobaya', further described in Appendices~\ref{sec::appendixA} and~\ref{section:appendixC}.

\subsection{Noise Realizations of DES-Y1 data vectors}
\label{sec:lcdm_data}
The DES-Y1 covariance matrix for cosmic-shear, galaxy-galaxy lensing, and galaxy clustering and the noiseless fiducial data vector are evaluated at the DES-Y1 best-fit cosmology using \verb'CosmoLike'. We use the DES-Y1 covariance matrix to generate hundreds of millions of (Gaussian) noise realizations around the noiseless fiducial DES-Y1 $\Lambda$CDM best-fit data vector. The generation of a large sample of noise realizations densely populates the $\vec \chi^2 = (\chi^2_{\text{shear}}, \chi^2_{\text{2x2}})$ space around our fiducial data vector. We then applied Kernel Density Estimator (KDE) to define, from the samples, confidence intervals of agreement. Based on these confidence regions we select 68 data vectors that lie at the $68\%$ (one $\sigma$), $99.7\%$ (three $\sigma$), and $99.99997\%$ (five $\sigma$) confidence intervals with approximate angular uniformity in $\vec \chi^2$ space.

The KDE method, implemented with help of \verb'GetDist'~\citep{Lewis:2019xzd} routines, approximates the probability distribution of a continuum of values for $\vec \chi^2$ from N generated samples $\vec \chi^2_{i=1,\cdot\cdot\cdot,\text{N}}$ as follows
\begin{align}
P(\vec{\chi}^2) = \sum_{i=1}^N K_f(\vec{\chi}^2-\vec{\chi}_i^2) \,
\end{align}
where $K_f$ is a multivariate Gaussian kernel with zero mean and covariance $f \times \hat{C}$ where $\hat{C}$ is the sample covariance of the $\vec \chi^2$. We found that given our large sample of computed data vectors $f \sim 0.1$ is a good choice to balance smoothing and noise features in the $P(\vec{\chi}^2)$ contours. Figure~\ref{fig:evidence_ratio1} shows the final selection of data vectors as seen in $\vec \chi^2$ space and displays the 1-5 $\sigma$ confidence intervals as determined by our selected KDE. The angular distribution of the selected noise realizations nicely covers all quadrants. Figure~\ref{fig:evidence_ratio1} also illustrates the evidence ratios of the selected data vector realizations, specifically the color bar shows the natural-log ratio of the data vector's 3x2pt evidence to its 2x2pt and shear evidences as defined in Eq.~\ref{eqn:evidence_ratio}.

\subsection{Simulated analysis of noisy data vector realizations}
\quad \quad Using the data vectors as generated in Sect.~\ref{sec:lcdm_data}, we now investigate whether  statistical fluctuations in the DES-Y1 data vector have a high probability of causing tension (as defined by the Jeffreys scale). 

Figure~\ref{fig:evidence_ratio1} shows that there is no radial or angular dependency in the value of the evidence ratio as a function of $\chi^2$ values in cosmic shear and 2x2. Similarly, Fig.~\ref{fig:evidence_ratio_hist} shows no differences in the evidence ratio distribution associated with one, three, and five $\sigma$ noise realizations; the histograms of evidence ratios are all centered on large positive values as predicted by~\citep{Raveri:2018wln} and~\citep{Handley:2019wlz} for wide uninformative priors. 

The comparison between the evidence ratio and suspiciousness (c.f. Fig. \ref{fig:evidence_ratio2}) shows that broad priors significantly increase the number of noise fluctuations that are not flagged as internal tension by evidence ratios, but they would be flagged by using suspiciousness. 
It is however not clear that a prior independent metric, such as suspiciousness, is necessarily more objective. While Bayesian evidence tends to hide tensions if broad priors are chosen, it is important to note that tensions in data are inevitably connected to our prior understanding of the situation.  \cite{Handley:2019wlz} argue that some known tensions in cosmology would have been interpreted differently had they been observed decades ago, when our prior beliefs encompassed a broader range. 

\begin{figure}
\centering
\includegraphics[width=0.475\textwidth,height=0.34\textwidth]{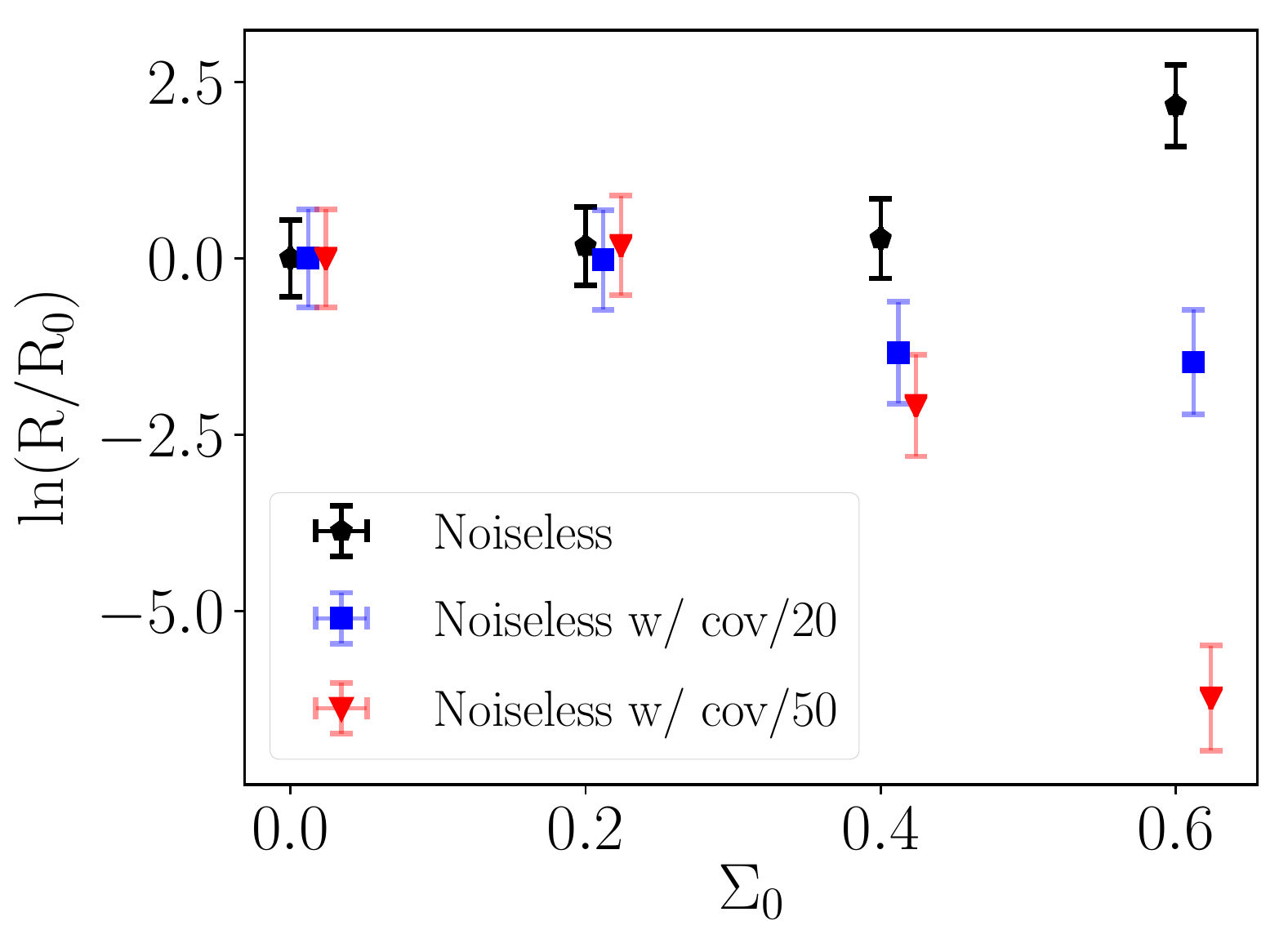}
\cprotect\caption{Comparison between the evidence Ratio, $R$, for models with $\Sigma_0 = \{0.2, 0.4, 0.6\}$ and the evidence, $R_0$, for the $\Lambda$CDM model ($\Sigma_0 = 0$) model. Black diamonds are chains with DES-Y1 covariance, while blue squares and red triangles are chains with covariances that were divided by 20 and 50, respectively. For DES-Y1 chains, the posterior for many parameters are being pressed against the prior boundaries before inconsistencies between cosmic shear and 2x2pt become important, which explains the unexpected behavior of evidence ratio going up as a function of $\Sigma_0$. 
}
\label{fig:mg_evidence}
\end{figure} 

It is difficult to estimate which tension estimator is a better choice. In Fig. ~\ref{fig:evidence_ratio2} (right panel), we present a comparison and relative calibration between evidence ratios and suspiciousness (for the specific DES-Y1 case considered in this paper). Our results show how metrics that rely, at least for Gaussian Likelihoods, solely on the likelihood of the data differ from tension estimators that take the DES-Y1 prior beliefs into account.

\begin{figure*}
\centering
\includegraphics[width=0.885\textwidth,height=0.885\textwidth]{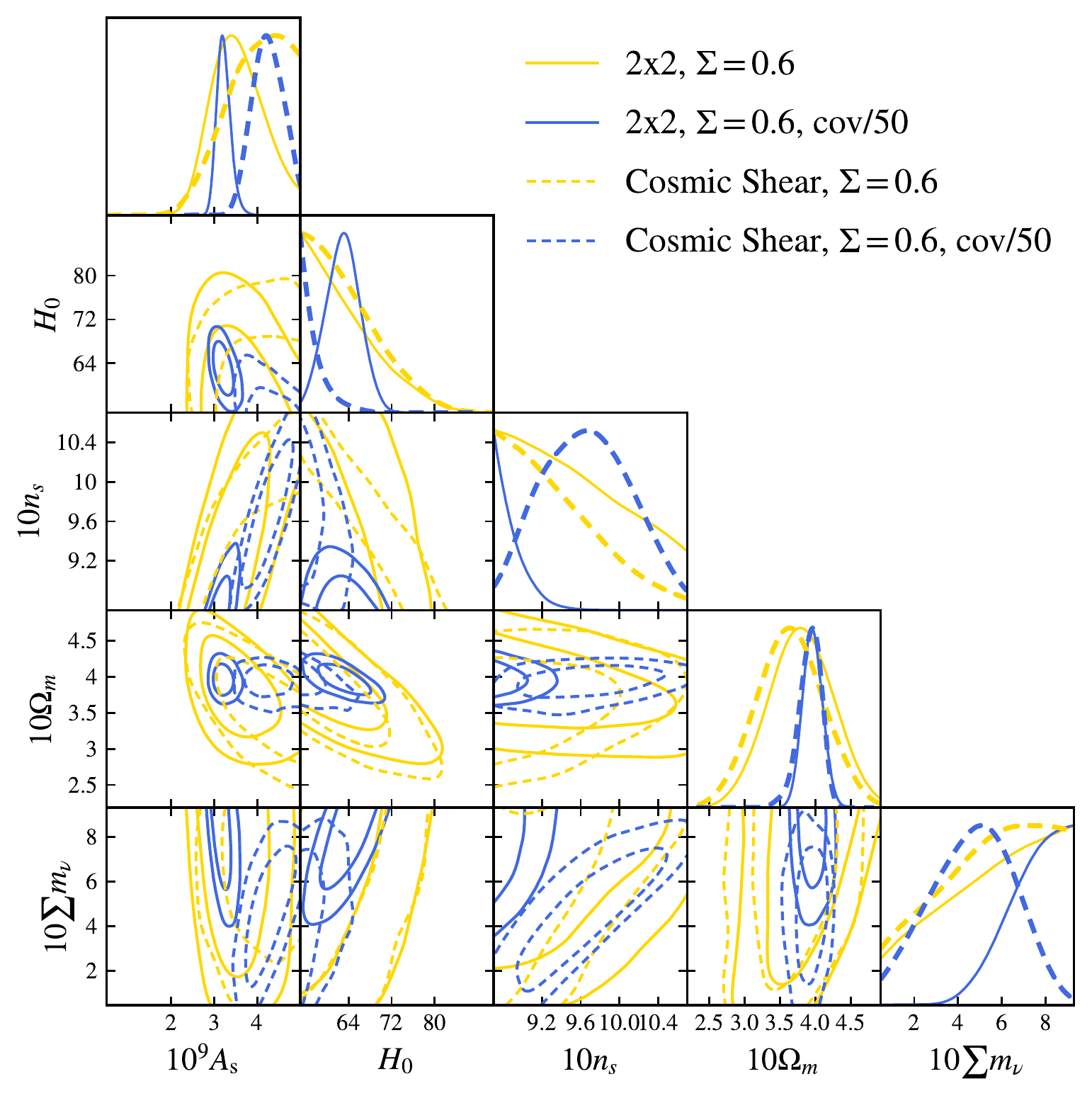}
\cprotect\caption{The posterior distribution of selected parameters for cosmic shear (dashed) and 2x2pt (solid) analyses, and for the default DES-Y1 covariance (yellow) against the case where the covariance was reduced by a factor 50 (blue). While it is true that $\Sigma_0 \neq 0$ predicts inconsistencies between the cosmological parameters in $\Lambda$CDM, it is difficult to see them in DES-Y1 chains. Not only the error bars are larger in DES-Y1, but also the posteriors are being squeezed against the prior boundaries.}
\label{fig:mg_posterior}
\end{figure*}

Figure \ref{fig:evidence_ratio_hist} shows that the observed DES-Y1 evidence ratio does not point towards an exceptional level of agreement between the datasets as would be inferred by the Jeffreys scale. Generally speaking we do not find a significant difference in the evidence ratio's mean or variance of data vectors drawn from the 1-$\sigma$, 3-$\sigma$, 5-$\sigma$ noise level (also c.f. Fig. \ref{fig:evidence_ratio2}, left panel). In addition, we also find that a noisy DES-Y1 data realizations from the 1-$\sigma$ confidence region of the parameter covariance matrix can have a negative evidence ratio, which would point towards a significant discrepancy. These findings make it difficult to motivate the DES-Y1 Bayesian evidence ratio as a strong indicator for significant agreement between cosmic shear and 2x2. 

In the case of correlated Gaussians, the evidence ratio and $\Delta \chi^2 = \chi^2_{12} -\chi^2_{1}-\chi^2_{2} $ (i.e. the maximum log-likelihoods) are linearly correlated. In our DES-Y1 posteriors, we however find that a linear combination of the log-likelihoods, defined as $\Delta \bar{\chi}^2$ (Eq. ~\ref{subsec::delta-chi-eq}), is correlated with the evidence ratio. No correlation was found when comparing evidence ratios against generalized parameter distances. 
	
\section{Evidence ratios with internal tension}	 
\label{sec:mg_noiseless}
In this section we investigate the evidence ratio's behavior when assuming a $\mu$-$\Sigma$ modified gravity scenario (as studied in \cite{Abbott:2018xao}, \cite{planck_2015_mu_sigma}, \cite{Aghanim:2018eyx}, and \cite{Simpson_2012}) that induces tension between the weak lensing and the galaxy clustering parts of the 3x2 data vector. Recall that $\Sigma \neq 0$ only affects cosmic-shear and galaxy-galaxy lensing. 
 
\subsection{Modified Gravity Data Vectors}
\label{sec:mg_data}
\quad\quad Following the definitions in~\cite{Ferreira_2010}, the Poisson and lensing equations in Newtonian Gauge are altered in the $\mu$-$\Sigma$ model as: 
\begin{align}
k^2 \Psi = -4 \pi G a^2 (1 + \mu(a)) \rho \delta \\
k^2 (\Psi + \Phi) = -8 \pi G a^2 (1+ \Sigma(a)) \rho \delta.
\end{align}

Similar to the $\Lambda$CDM case (c.f. Sect. \ref{sec:lcdm_data}), we compute the $\mu$-$\Sigma$ data vector at the DES-Y1 best fit parameter values. Specifically, we set $\mu(a) = \mu_0 \, \Omega_\Lambda(z)/\Omega_\Lambda$ and $\Sigma(a)= \Sigma_0 \, \Omega_\Lambda(z)/\Omega_\Lambda$, with $\Omega_\Lambda(z)$ being the redshift dependent dark energy density over the critical density. No noise is added to the modified gravity data vectors. Similar to the $\Lambda$CDM cases, we apply \verb'Halofit' \citep{Takahashi:2012em} to compute the nonlinear matter power spectrum in the $\mu-\Sigma$ case. The fact that \verb'Halofit' does not correctly describe the nonlinear physics of $\mu-\Sigma$ gravity is not a significant concern for this paper since it is not out goal to analyze actual data. Instead our goal is to examine changes in the evidence ratio when the data vector is computed from a different underlying physics than the model that is assumed in the analysis.

\subsection{Simulated Likelihood Analysis - modified gravity induced tension}
\label{sec:mg_like}
\quad \quad We now investigate induced internal tensions in the case where a data vector originating from $\mu$-$\Sigma$ gravity (see Sect.~\ref{sec:mg_data} for definitions) is evaluated in the DES-Y1 pipeline for a $\Lambda$CDM cosmology. We have generated fiducial data vectors with fixed $\mu = 0$ and $\Sigma$ ranging from $0 \leq \Sigma_0 \leq 1$. We have not added noise realizations from DES-Y1 covariances; the modified gravity data vector is noise free. Figure~\ref{fig:mg_evidence} presents a surprising behavior of evidence ratios: the log-evidence ratio of the noiseless  modified gravity data vector and our fiducial noiseless $\Lambda$CDM data vector increases as a function of $\Sigma_0$ (black diamonds). This means that the physical tension introduced by the modified gravity parameters in the galaxy clustering, galaxy-galaxy lensing, and cosmic shear parts of the data vector is not identified as such by the Bayesian evidence ratio. 

Such unexpected behavior of the evidence ratio can be better understood by looking at Fig. ~\ref{fig:mg_posterior}. We see that several parameters are pushing against the prior boundaries. This boundary effect reduces differences between the cosmological parameters that fit cosmic shear and 2x2pt at the expense of making the goodness of fit between theory and data worse. To check that prior boundaries are indeed responsible for the unusual behavior of the evidence ratio, we re-examine the log-evidence ratio of the noiseless  modified gravity data vector and our fiducial noiseless $\Lambda$CDM data vector, however this time we rescale the covariance matrices by factors of twenty (c.f. Fig. \ref{fig:mg_evidence} blue squares) and fifty (c.f. Fig. \ref{fig:mg_evidence} red triangles). This rescaling procedure significantly reduces the posterior volume, which reduces or even removes the prior boundary effects. Indeed, the evidence ratio now decreases as a function of $\Sigma_0$ as expected. This type of behavior exemplifies the difficulties in interpreting tension metrics in realistic examples without extensive validation via simulated analyses.

\section{Conclusion}
\label{sec:conclusion}
\quad \quad Tension metrics are an important aspect of multi-probe analyses; they will be used increasingly to determine whether probes can be combined or whether tension across probes need to be further explored. However, tension metrics themselves need to be calibrated by simulated analyses for each dataset in order to define levels of discordance.

In this work we study the properties of several tension metrics for the specific case of the DES-Y1 3x2pt analysis. In \cite{Abbott:2017wau} the individual analyses of 1) cosmic shear and 2) the galaxy-galaxy lensing plus galaxy clustering (so-called 2x2pt) were compared and ultimately combined into a so-called 3x2pt analysis. Both data vectors, cosmic shear and 2x2pt, were deemed consistent under an assumed $\Lambda$CDM model. Consistency was demonstrated by computing the Bayesian evidence ratio, with the result of 6.39, and interpreted using the Jeffreys scale. Bayesian evidence ratios however are known to be prior dependent and it is important to calibrate the computed numbers through a large suite of simulated analyses. 

In this paper we calibrate the distribution of evidence ratios for a large set of noise realizations around the DES-Y1 best fit $\Lambda$CDM cosmology. The noisy data vectors are drawn from the DES-Y1 data covariance, not from the parameter covariance. While the data covariance and parameter covariance are closely related, noise realizations drawn from the low-dimensional parameter covariance map onto smooth modulations in the 457-dimensional data space with little scatter from the fiducial data vector. Our data covariance includes Gaussian cosmic variance, shot/shape noise (for clustering/weak lensing, respectively), and non-Gaussian contributions to the covariance from the connected four-point function of the matter density field as well as super-sample covariance (SSC)~\citep{tah13}. As the Gaussian cosmic variance terms and shape/shot noise are caused, respectively, by the limited number of independent Fourier modes sampled in each angular bin and the limited number of galaxies sampled in the power spectrum measurement, noise realizations drawn from the data covariance are nearly uncorrelated between different Fourier modes and provide "noisy" scatter with little noticeable bias from the fiducial data vector." 

We run multiple simulated likelihood analyses for a DES-Y1 cosmic shear, 2x2pt, and 3x2pt data vector and find that the Bayesian evidence value obtained by DES-Y1 (6.39) is rather typical.  We then explore evidence ratios where noiseless data vectors that are computed from a $\mu-\Sigma$ modified gravity model are analyzed with a pipeline that assumes a $\Lambda$CDM model. Under these assumptions, a physical tension is induced between the weak lensing and galaxy clustering parts of the 3x2pt data vector and we explore the Bayesian evidence ratio behavior as a function of increasing the strength of the modified gravity model (increasing $\Sigma$).
We demonstrate that prior boundary effects can efficiently hide tensions between the weak lensing and galaxy clustering part of the 3x2pt data vector. When significantly increasing the constraining power, by dividing the covariance by factors 20 and 50, we show that such boundary effects are significantly reduced and the expected tension appears.

Our findings confirm that the evidence ratio, as measured by the Jeffreys scale, is biased towards compatibility between the datasets due to DES-Y1's adopted priors. These wide priors were intentionally chosen conservatively and did not take into account prior knowledge from other experiments. Such wide priors have the potential to hide tensions between probes. In the near future DES data quality will be superseded by stage IV experiments, in particular, Rubin Observatory's LSST (\cite{lsst}), SPHEREx (\cite{spherex}), Euclid (\cite{euclid}), and the Roman Space Telescope (\cite{wfirst_spergel}, \cite{wfirst_eifler}). These experiments will provide an unprecedented amount of high-quality data that will enable not just 3x2pt analyses, as considered in this paper, but a large variety of other cosmological probes as well. Exploring tensions between probes of the same data set and (even more interesting) between datasets will be a critical part of the data analysis of these missions, throughout which simulated analyses to calibrate tension metrics should become a standard tool in precision cosmology.  

\section*{Acknowledgments}
We want to thank T. Eifler for the thorough review and extensive suggestions that improved our results' presentation. We would also like to thank P. Lemos and M. Raveri for fruitful discussions. VM is supported by NASA ROSES 16-ADAP16-0116 and NASA ROSES ATP 16-ATP16-0084. PR and EK are supported by Department of Energy grant DE-SC0020247.
\appendix
\section{Pipeline Validation}
\label{sec::appendixA}
\begin{figure*}
\centering
\includegraphics[width=0.325\textwidth,height=0.25\textwidth]{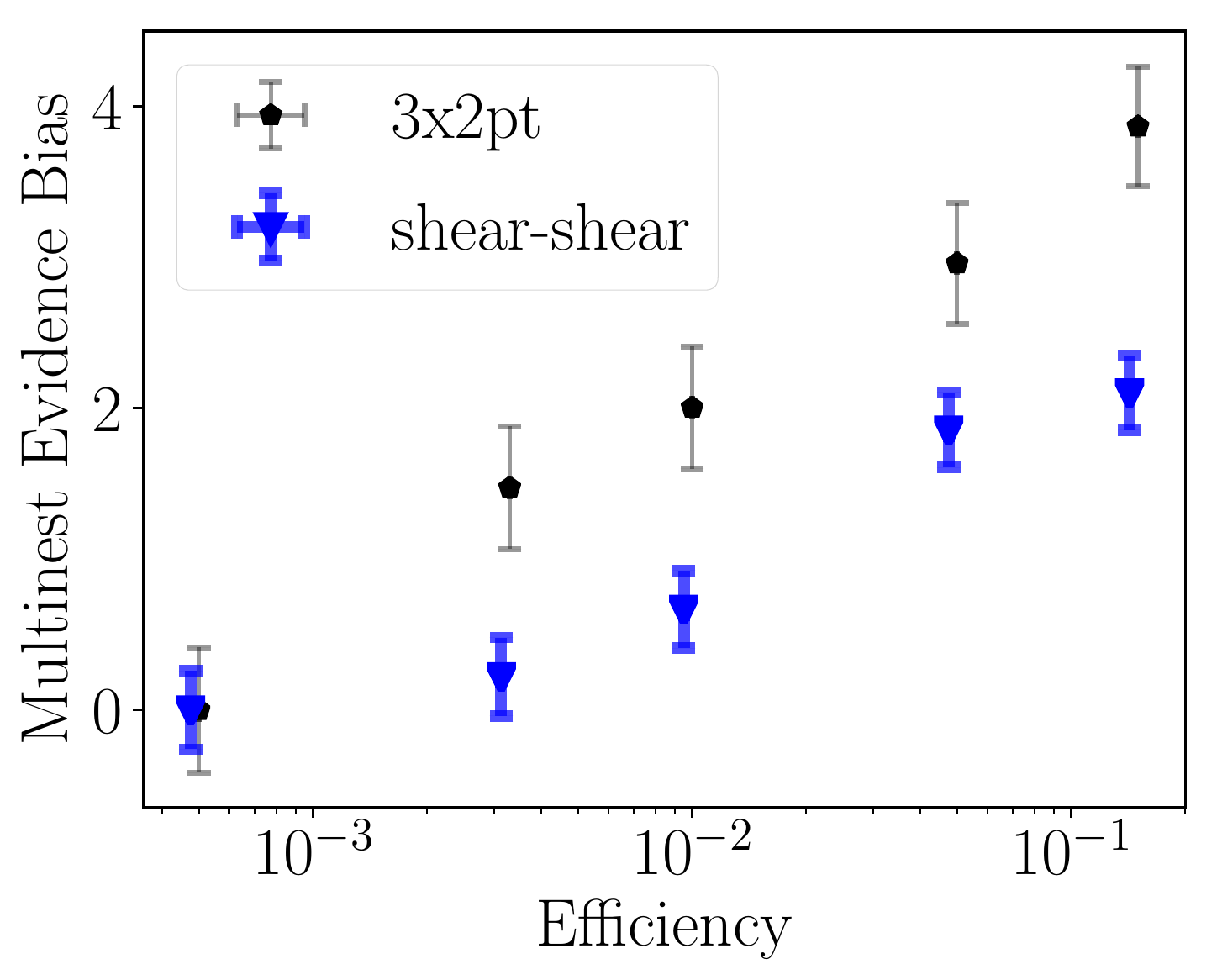}
\includegraphics[width=0.325\textwidth,height=0.25\textwidth]{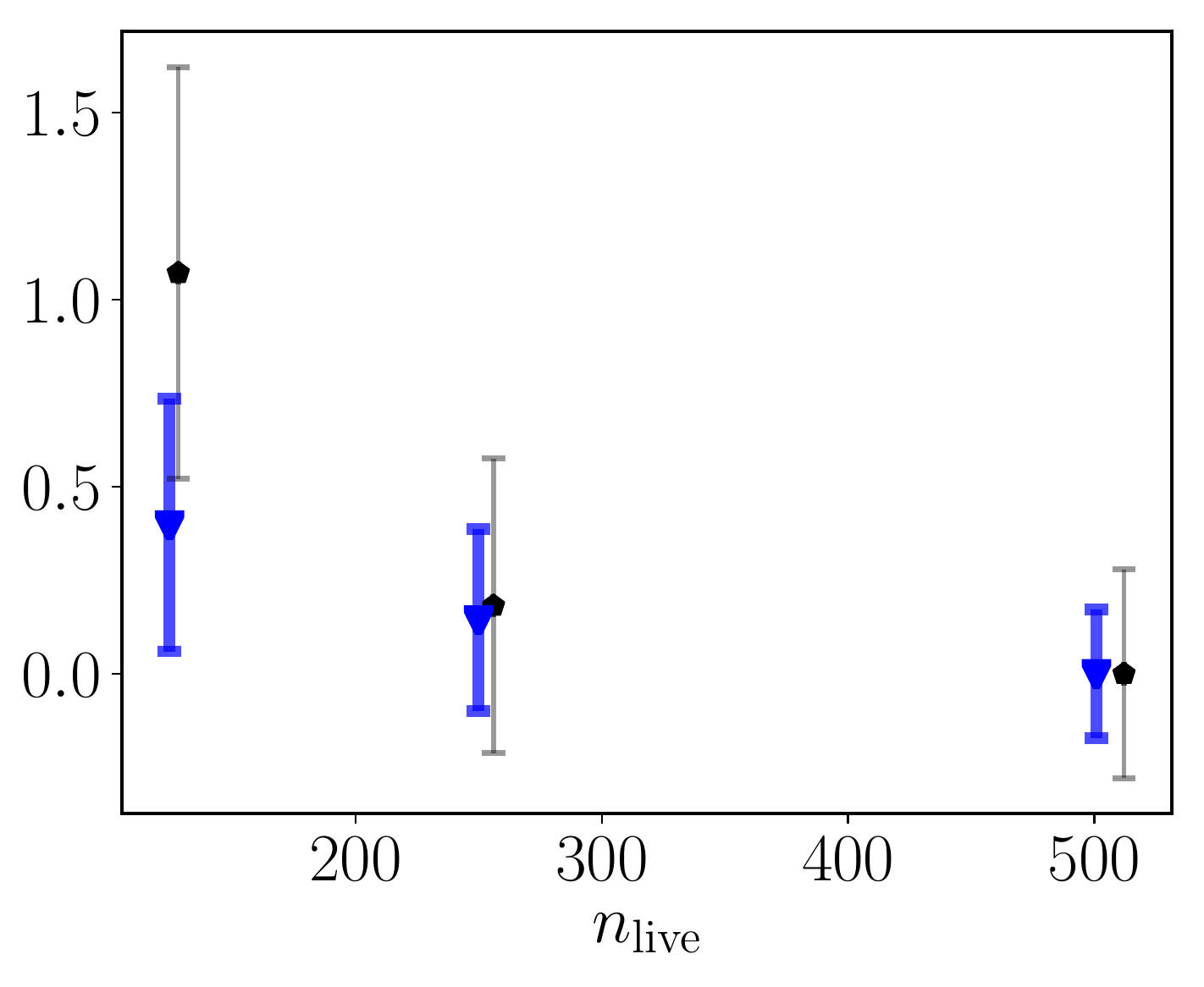}
\includegraphics[width=0.325\textwidth,height=0.25\textwidth]{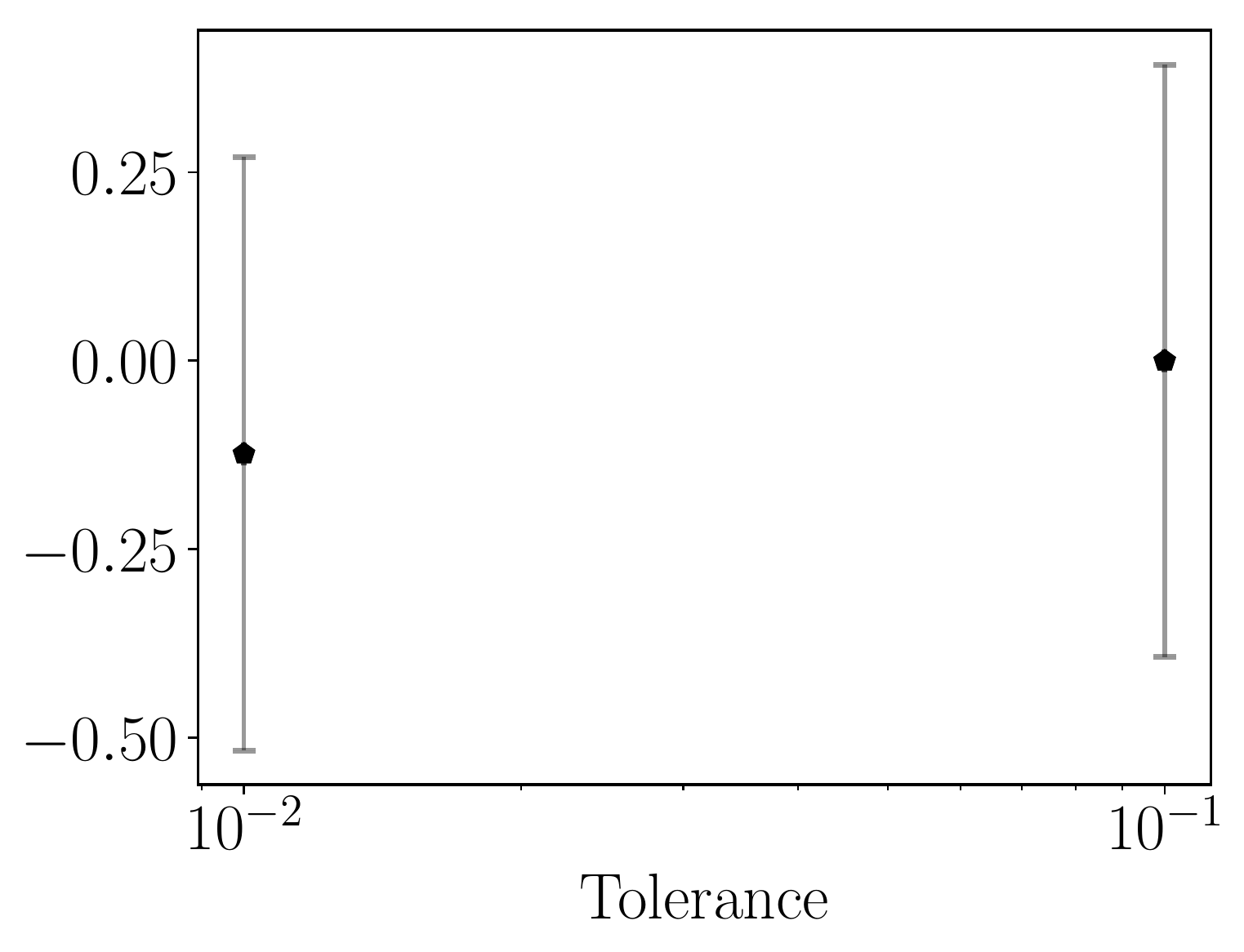}
\cprotect\caption{\verb'MultiNest' evidence bias as a function of the sampling efficiency (left panel),  number of live points (middle panel) and evidence tolerance factor (right panel). As a simplifying assumption, the evidence evaluated from the chain with either the lowest efficiency or the highest number of live points or the tolerance factor has zero bias by construction. The error bars reflect \verb'MultiNest''s claimed uncertainties and no error propagation was applied to take into account the error bars in the value of the unbiased evidence.}
\label{fig:mn_calib}
\end{figure*} 

\begin{table*}
\centering
\begin{tabular}{ l  c c  c c c c | c c  }
\hline Sampler   & $3\times2$pt \verb+DV0+   & $3\times2$pt \verb+DV1+ & $2\times2$pt \verb+DV0+ & $2\times2$pt \verb+DV1+ & cosmic shear \verb+DV0+  & cosmic shear \verb+DV1+ & R \verb+DV0+ & R \verb+DV1+  \\ \hline
GLM - Mean & -306.4 & -204.0 &  -172.4 & -116.3 & -154.5 & -110.89 & 20.5 & 23.2  \\
GLM - Chain BF & -307.5 & -204.6 &  -176.4 & -117.7 & -142.1 & -91.7 & 11 & 4.8 \\
GLM - MKL & -306.4 & -204.6  &  -176.4 &  -117.7  &  -154.5  & -110.89& 24.5 & 23.9 \\
Polychord &  $-307.1 \pm   0.4$ & $-204.8 \pm  0.4$  & $-171.8 \pm   0.4$ & $-117.4 \pm  0.4$ & $-143.2\pm   0.3$  & $-94.8 \pm   0.3$ & 7.9  & 7.4 \\
\end{tabular}
\cprotect\caption{The Comparison performed between predicted Bayesian evidence evaluated using \verb'MultiNest', \verb'PolyChord' and Gaussian Linear Modeling of Metropolis-Hasting chains around either the median of the parameters or the chain best fit. MKL stands for Minimum Kullback-Leibler divergence~\citep{kullback1951}, and in that row, we select the Gaussian approximation from the two previous cases by minimizing the KL divergence against the full posterior~\citep{Raveri:2018wln}). In all cases, the  additional constraint $0.005 < \Omega_\text{b} h^2 < 0.04$ were applied as an additional top-hat likelihood. \verb'DV0' and \verb'DV1' represent distinct noise realizations of the best-fit data vector.}
\label{tab:gaussian_test_models}
\end{table*}

\begin{figure}
\centering
\includegraphics[width=0.475\textwidth,height=0.475\textwidth]{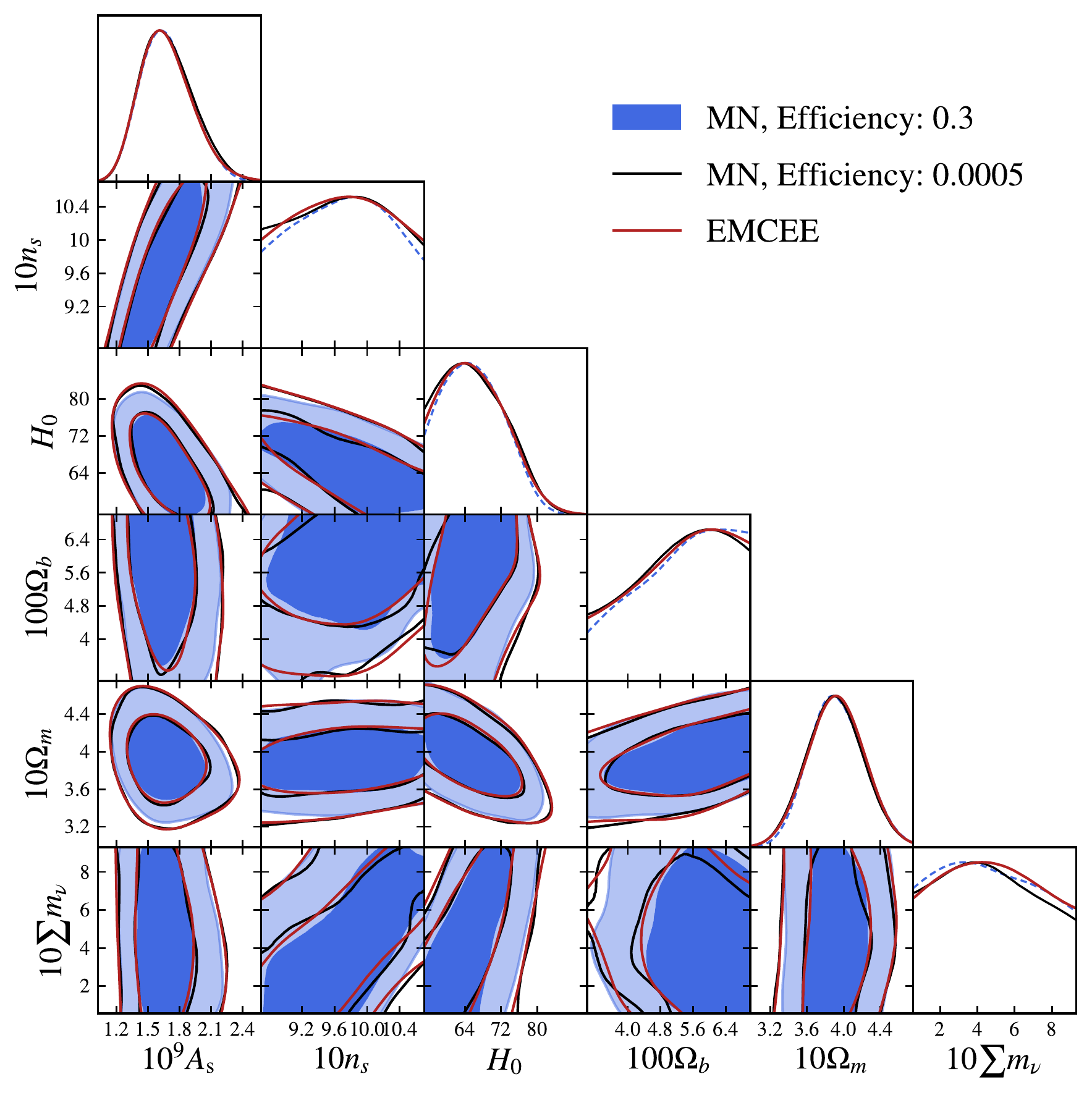}
\cprotect\caption{The panel presents the posterior predicted by \verb'Multinest' as a function of the adopted efficiency hyperparameter. Table~\ref{tab:models} shows the values of additional \verb'Multinest'  settings. The comparison against the \verb+Emcee+ sampler confirms that chains with high-efficiency do predict posteriors that are quite close to the truth. Indeed, no posterior feature stands out as being an outlier, something that would indicate that lower efficiency is indeed needed as it predicts order unity bias for the evidence (see Figure~\ref{fig:mn_calib}).}
\label{fig:mn_calib_posterior}
\end{figure}

\begin{table}
\centering
\begin{tabular}{ l  c c  c  c }
\hline Sampler   & $n_{\text{live}}$  & Efficiency & Tolerance & $n_{\text{repeats}}$ \\ \hline
Multinest (MN) & $256$ & $0.3$ & $0.1$ &  -- \\
Polychord & $256$ & --  & $0.05$  & $3\times\mathrm{dim}$  \\
\end{tabular}
\cprotect\caption{Default values assumed for the internal parameters employed in the multiple sampler codes we analyzed in our appendix. In regards to \verb'MultiNest', tolerance corresponds to the \textit{evidence tolerance factor}; efficiency is the \textit{sampling efficiency} (the variable \verb'efr') and $n_\text{live}$ matches the \textit{number of live points}. In addition, we set to \verb'False' the boolean variable that sets up the \textit{constant efficiency mode}. Using \verb'PolyChord', \textit{clustering} was turned off by default, and $n_\text{repeats}$ matches the variable \verb'num_repeats'. \verb+Emcee+ runs consume a fixed amount of computer resources to ensure that chains contain no less than 5 million samples. On the other hand, Metropolis Hasting samples were run until reaching convergence according to the Gelman and Rubin criteria, where we find the mean and standard deviation of the Gelman-Rubin criteria to be 0.02 and 0.2, respectively.}
\label{tab:models}
\end{table}

\begin{figure}
\centering
\includegraphics[width=0.475\textwidth,height=0.475\textwidth]{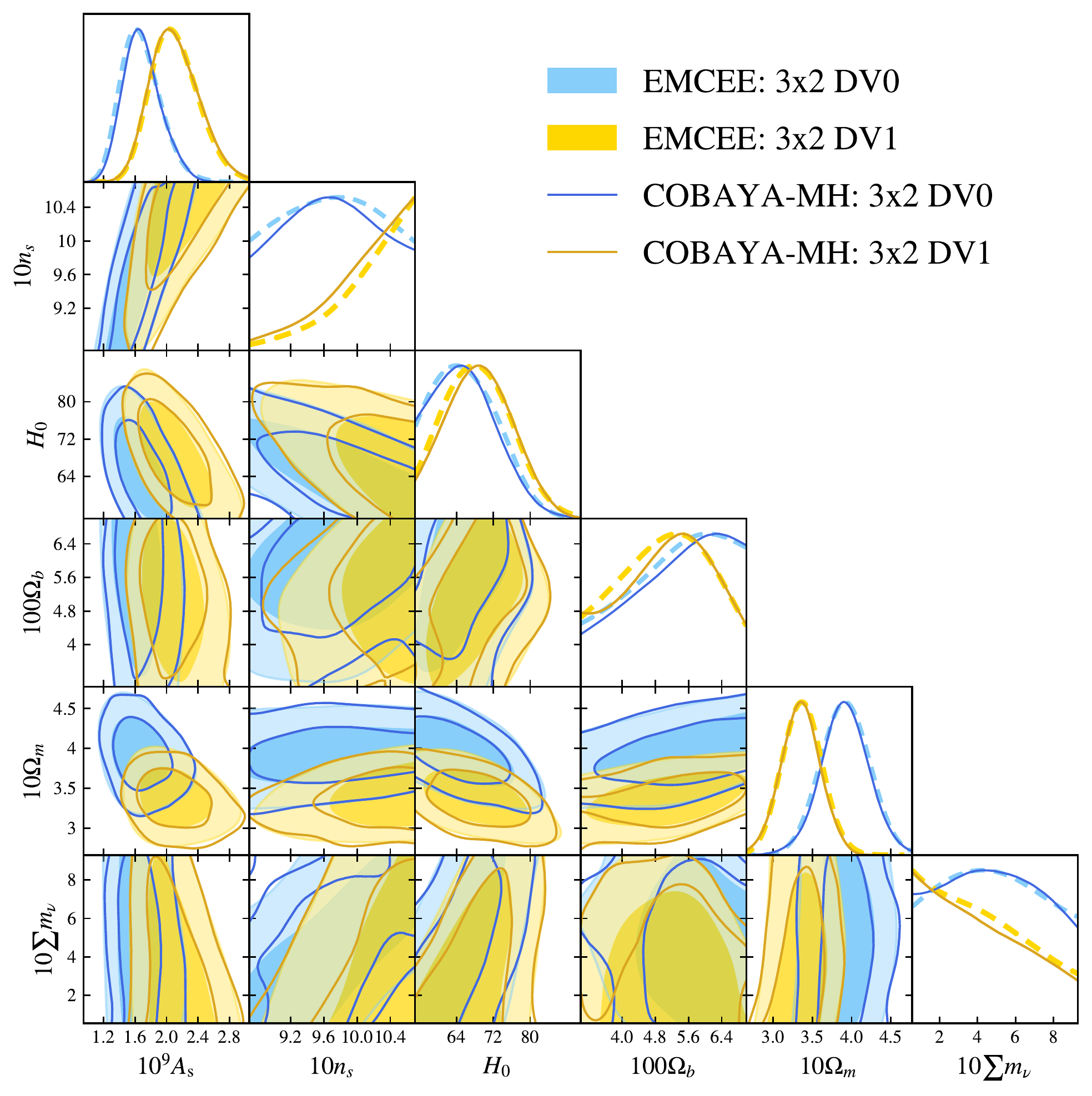}
\cprotect\caption{The figure compares the predicted posterior for the cosmological parameters given by \verb+Emcee+ and Metropolis-Hasting samplers. Blue shades on the two-dimensional panels correspond to dashed blue lines on the 1D posterior plots. The two 3x2pt data vectors - \verb'DV0' and \verb'DV1' were data vectors with noise generated using a simulated DES-Y3 covariance. The agreement between the two samplers is good to cross-check, considering the pipelines are somewhat different: the linear power spectrum on \verb+Emcee+ was evaluated within \verb+CLASS+ (default \verb'CosmoLike' pipeline) while for the Metropolis-Hasting we have performed a merging between \verb'Cobaya' and \verb'CosmoLike' and used \verb'CAMB' to calculate the matter power spectrum.}
\label{fig:emcee_vs_MH_posterior}
\end{figure}

\begin{figure}
\centering
\includegraphics[width=0.475\textwidth,height=0.475\textwidth]{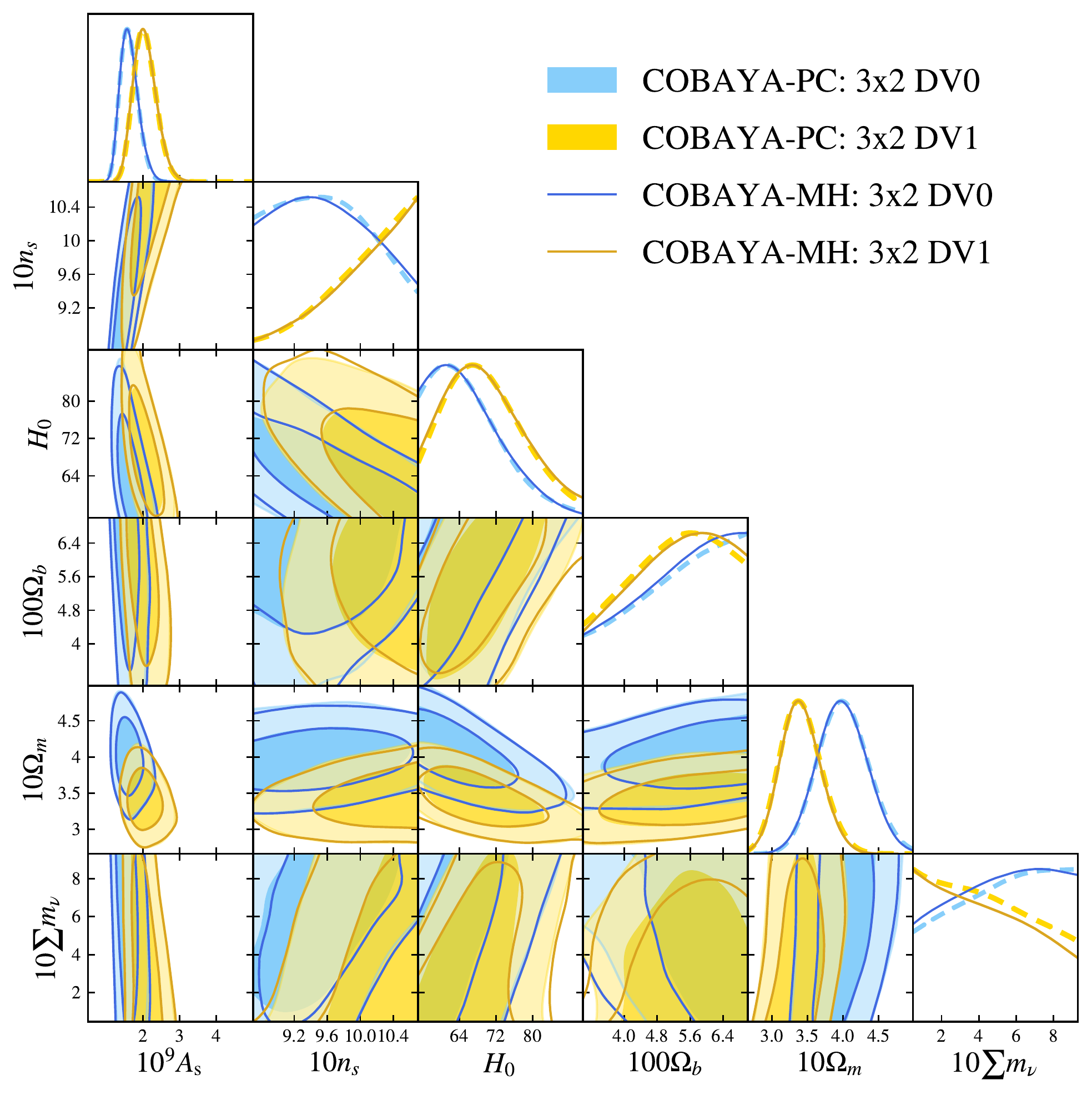}
\cprotect\caption{The figure compares the predicted posterior for the cosmological parameters given by \verb'Polychord' against Metropolis-Hasting. Shades on the 2D panels correspond to dashed lines on the 1D posterior plots.  The two 3x2pt data vectors - \verb'DV0' and \verb'DV1' were data vectors with noise generated using a simulated DES-Y3 covariance. In both cases, the matter power spectrum was evaluated using \verb+CAMB+ (without removing the extra \verb+Halofit+ factor shown in Eq. ~\ref{eqn:halofit_extra_term}).}
\label{fig:poly_vs_MH_posterior}
\end{figure}

\begin{figure}
\centering
\includegraphics[width=0.475\textwidth,height=0.455\textwidth]{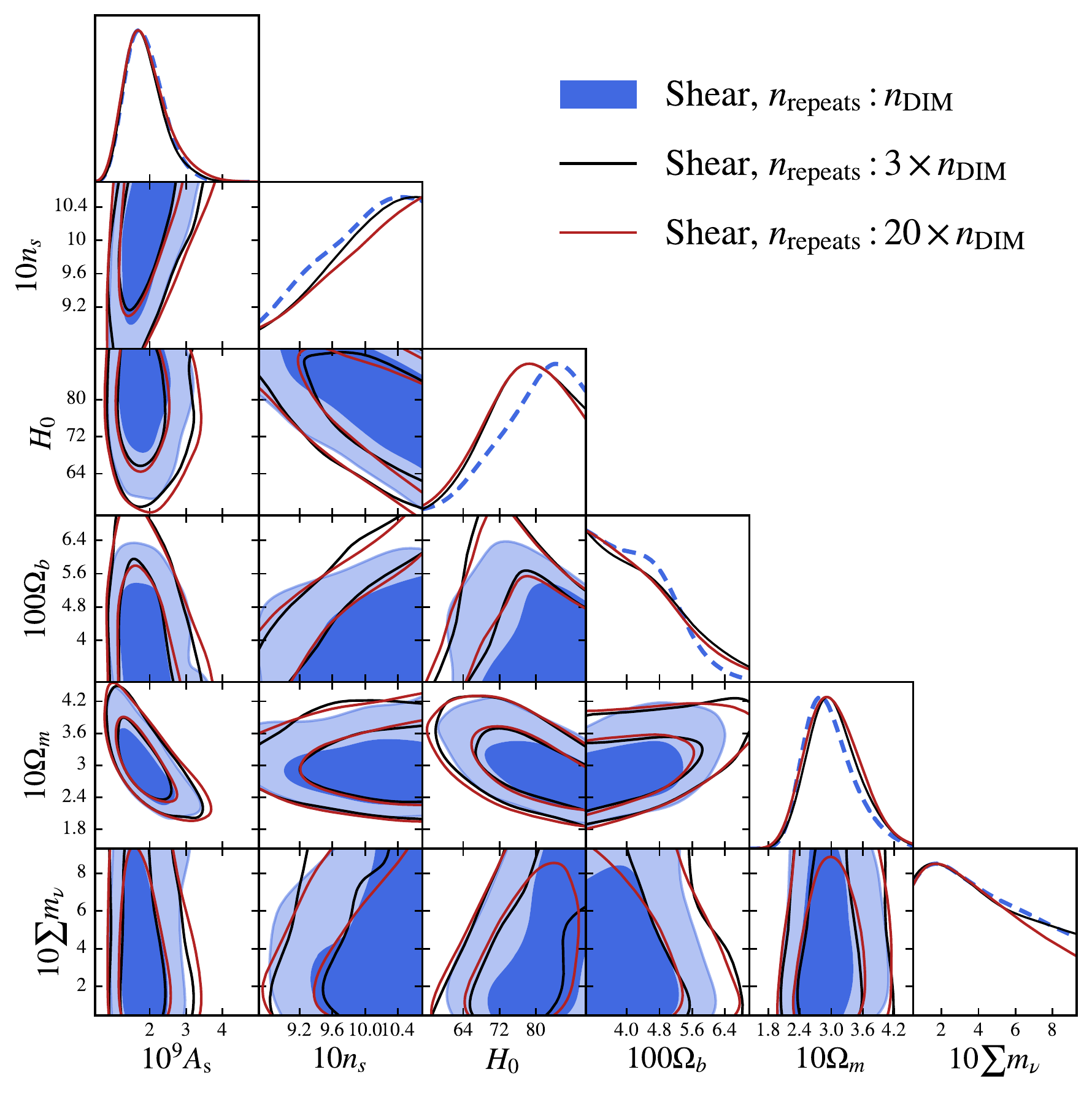}
\cprotect\caption{The figure compares the predicted posterior for the cosmological parameter given by \verb'Polychord' as a function of the hyperparameter $n_\mathrm{repeats}$  written in units of the number of parameters in the chain ($n_\mathrm{DIM}$). Blue shades on the two-dimensional panels correspond to dashed blue lines on the 1D posterior plots. On shear-shear, the posterior shows uncertain behavior in the case $n_\mathrm{repeats} = n_\mathrm{DIM}$, with no appreciable changes were seen in the range  $3 < n_\mathrm{repeats}/n_\mathrm{DIM} < 20$. This is not necessarily the case for 3x2pt data vectors, where setting $n_\mathrm{repeats} = n_\mathrm{DIM}$ is acceptable for posteriors. 
}
\label{fig:poly_numrepeat_posterior}
\end{figure}

\begin{figure*}
\centering
\includegraphics[width=0.325\textwidth,height=0.25\textwidth]{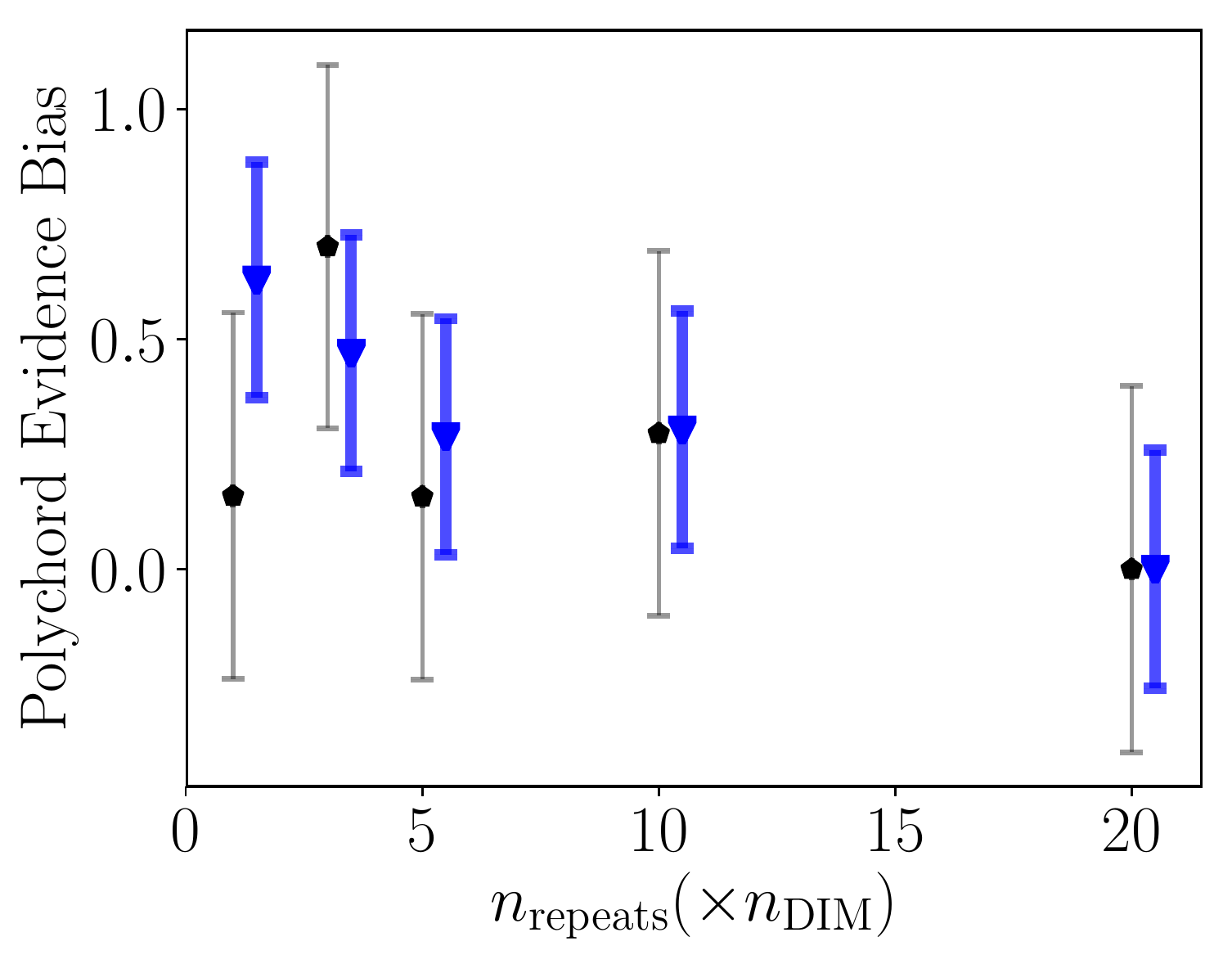}
\includegraphics[width=0.325\textwidth,height=0.25\textwidth]{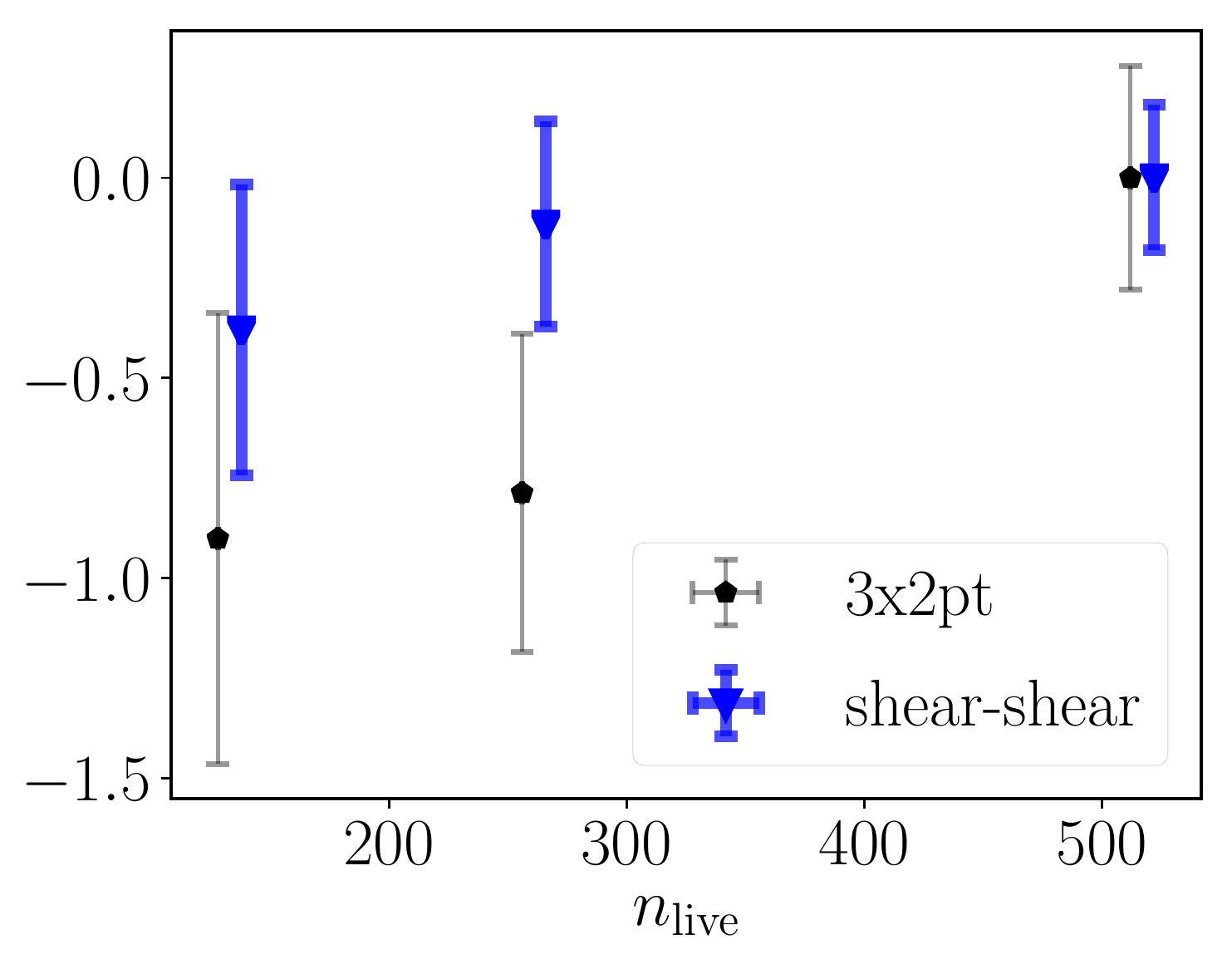}
\includegraphics[width=0.325\textwidth,height=0.25\textwidth]{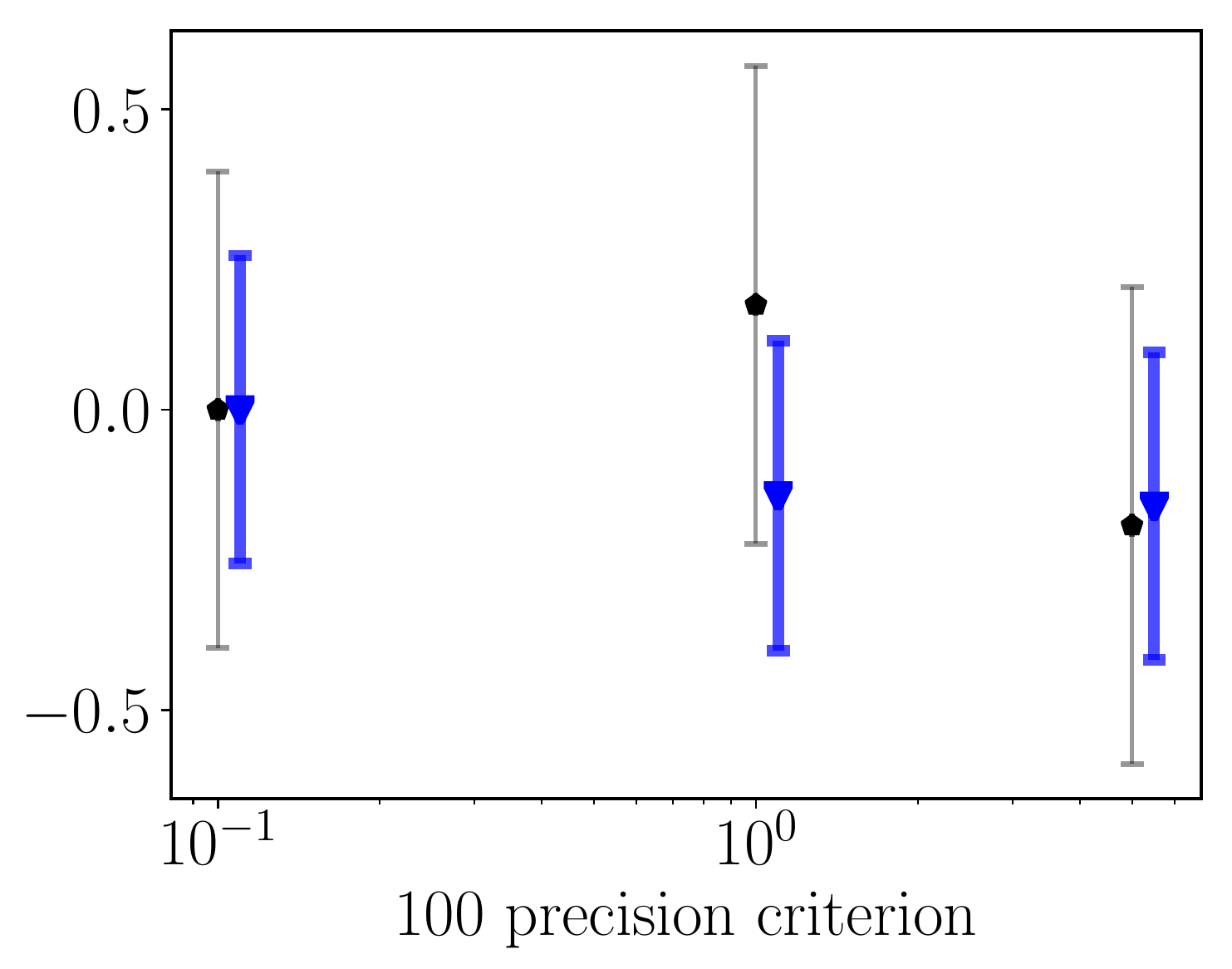}
\cprotect\caption{\verb'Polychord' evidence bias as a function of the $n_\text{repeats}$ parameter (left panel),  number of live points (middle panel) and precision criterion (right panel). As a simplifying assumption, the evidence evaluated from the chain with the highest $n_\text{repeats}$ (left panel), the highest number of live points (middle panel), or the lowest precision criterion factor (right panel) has zero bias by construction. The parameter $n_\text{repeats}$ on the left panel is shown in units of the parameter dimension, $n_\mathrm{DIM}$. The error bars reflect \verb'Polychord''s claimed uncertainties, and no error propagation was applied to take into account the error bars in the value of the unbiased evidence. Computational costs scale as $O(n_\mathrm{repeats})$~\citep{2015MNRAS.453.4384H}, the main bottleneck of our chains,  so we have adopted $n_\mathrm{repeats} = 3 \times n_\mathrm{DIM}$ as a middle ground between accuracy and computational costs.}
\label{fig:poly_calib}
\end{figure*}

\quad \quad This appendix focuses on the technical aspects of the pipeline calibration. As shown in the main manuscript, the DES posteriors are non-Gaussian in some dimensions, while the DES priors are partially informative in several directions, where the likelihood is weakly constraining. Such properties affect the required calibration of samplers hyperparameters, such as the \verb'Multinest''s efficiency~\citep{Feroz:2013hea}, given that the entire volume of the parameter space needs to be well sampled. Indeed, regions in parameter space with low non-negligible likelihood probabilities can contribute to the Bayesian evidence as long as there is enough prior volume where the likelihood has similar values. 

The default \verb'Multinest' configuration on DES-Y1 is: number of live-points $n_\text{live}= 500$, tolerance $= 0.1$ and efficiency $=0.3$. Figure~\ref{fig:mn_calib} reveals biases in the evidence values with such settings. For other hyperparameters, such as the number of live-points, changes in the reported evidence are compatible with the quoted error bars. These statements are valid for both the shear-only and the 3x2pt analyses. One prominent feature on figure~\ref{fig:mn_calib} is the constant slope of the evidence bias as a function of the \verb'Multinest''s efficiency in the case of the 3x2pt analysis.  There is no guarantee, therefore, that even efficiencies of the order of $10^{-4}$ would provide reliable results, and such settings raise the evidence's computational costs by one order of magnitude in comparison to the hyperparameter values adopted on DES-Y1. We emphasize that no conclusions on the general applicability of \verb'Multinest' can be drawn from our analysis; results are specific to DES-Y1. Figure~\ref{fig:mn_calib} also does not imply that there are no settings where \verb'Multinest' provides unbiased evidence ratios.

We also checked if the detected biases on \verb'Multinest' reported evidences could have been identified through features in the posterior by-product, something that would have called the attention as being flagrantly corrupted. Figure~\ref{fig:mn_calib_posterior} shows no substantial deviations in the posterior as a function of the efficiency parameter, except for slight enlargement of the two sigma contours, and we have run similar chains using the \verb+Emcee+~\citep{emcee} sampler to confirm such statement.  Comparisons different \verb'Multinest' and \verb+Emcee+ require robust calibration on both samplers, as one could argue that direct comparison could point to problems in \verb+Emcee+. 

To double-check that convergence on \verb+Emcee+ has been achieved, we have run extremely long chains to check the consistency of our results. Also, we have compared on Figure~\ref{fig:emcee_vs_MH_posterior} \verb+Emcee+ against a third sampler - Metropolis-Hasting - where the well established and reliable Gelman-Rubin criteria~\citep{gelman1992} for convergence can be applied. Such comparison also cross-checks our code development, which unites \verb+Cosmolike+ and \verb+Cobaya+ pipelines\footnote{\url{https://github.com/CosmoLike/cocoa}}. In our new code, \verb+Cosmolike+ receives distances, parameter values and the matter power spectrum as function of redshift and wavenumber and returns the DES-Y1 data vector. This merging allowed us to use both \verb'Polychord' and Metropolis-Hasting samplers with the fast-slow decomposition commonly adopted in CMB analyses~\citep{2005math......2099N, Lewis:2013hha}, while \verb+Emcee+ and \verb'Multinest' chains employ the original standalone \verb+Cosmolike+.

It is unclear how much \verb+Multinest+'s biases might have affected DES-Y1 official results, and it is beyond the scope of this article to make such an in-depth analysis of the DES-Y1 official chains. We do, however, believe that \verb+Cobaya-Cosmolike+ code combines the pipeline validation effort that has been performed on \verb+Cosmolike+ with samplers that are more robust than \verb+Multinest+ in evaluating Bayesian evidence ratios. \verb+Cobaya-Cosmolike+ also provides Metropolis-Hasting with fast-slow decomposition that possesses robust convergence criteria, which is hard to be assessed in \verb+Emcee+. Indeed, the posterior comparison between Metropolis-Hasting and \verb+Polychord+ show perfect agreement, as seen in figure~\ref{fig:poly_vs_MH_posterior}. Moreover, Figures~\ref{fig:poly_numrepeat_posterior} and~\ref{fig:poly_calib} show that \verb+Polychord+'s evidence and posterior are robust against variations on  the adopted values for its hyperparameters.

One additional issue emerged from the comparison between \verb+CAMB+ and \verb+CLASS+ Boltzmann codes. While the original \verb+Cosmolike+ is directly integrated to \verb+CLASS+, \verb+Cobaya+ framework provided, at the time we run our simulations, full support only to \verb+CAMB+\footnote{ \url{https://github.com/CobayaSampler/cobaya/issues/46}.}. Differences between \verb+CAMB+ or \verb+CLASS+ should have been negligible, but we did detect an extra factor on the \verb+Halofit+ formula implemented by \verb+CLASS+. We then modified \verb+CAMB+ to match \verb+CLASS+ choices, and we discuss this issue in greater depth on appendix~\ref{section:appendixC}. In addition to that, \verb+CLASS+ has limitations on the $\Omega_b h^2$ range when dealing with BBN constraints and because of that  \verb+Cosmolike+ does assume the prior $0.005 < \Omega_b h^2 < 0.04$. We, therefore, applied the same prior choice in the \verb+Cobaya+-\verb+Cosmolike+ joint pipeline. We do not expect such minor choices to affect the qualitative conclusions of this work.

\section{Gaussian Approximation}
\label{section:apxB}
\begin{figure*}
\centering
\includegraphics[width=0.405\textwidth,height=.311\textwidth]{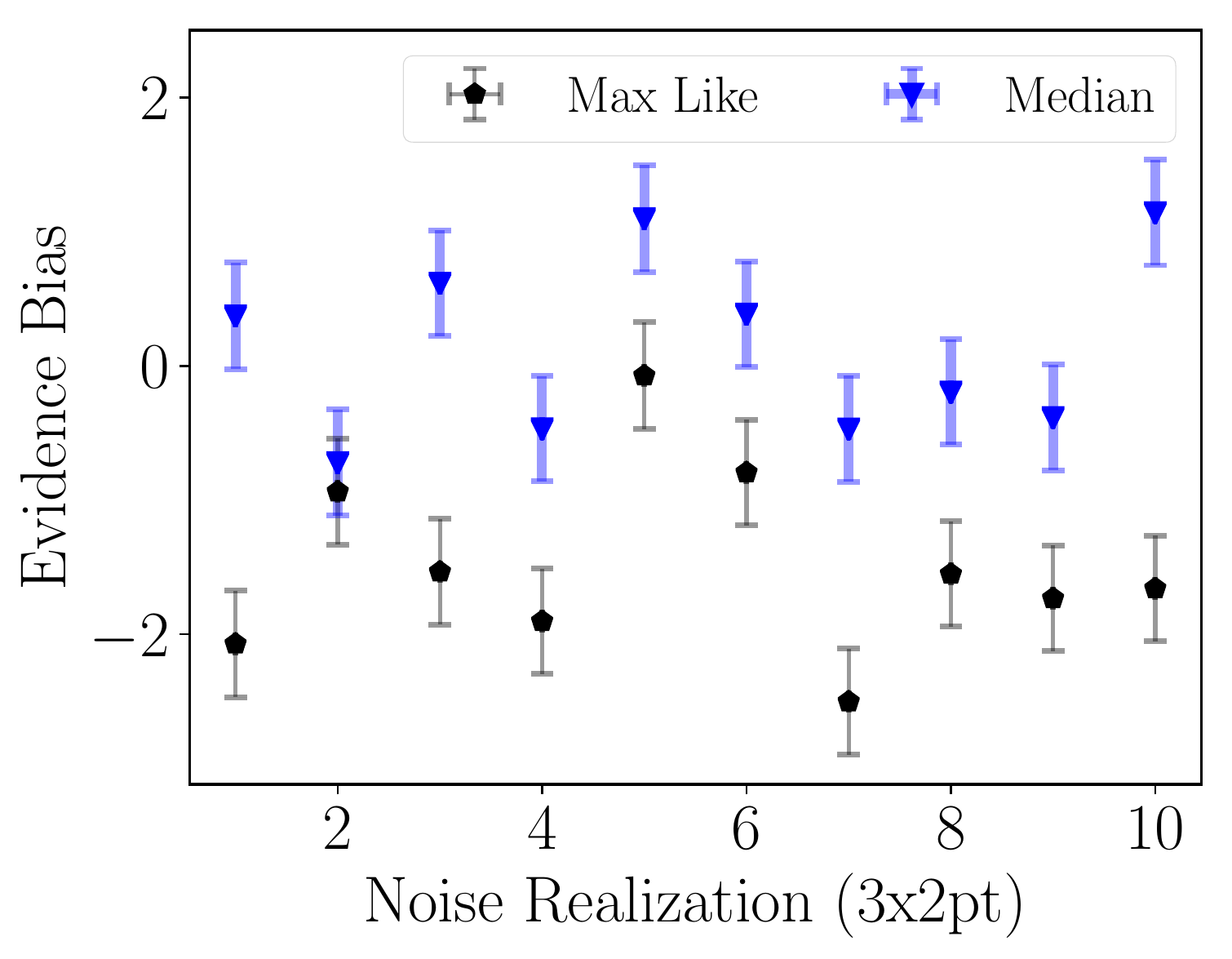}
\includegraphics[width=0.405\textwidth,height=.311\textwidth]{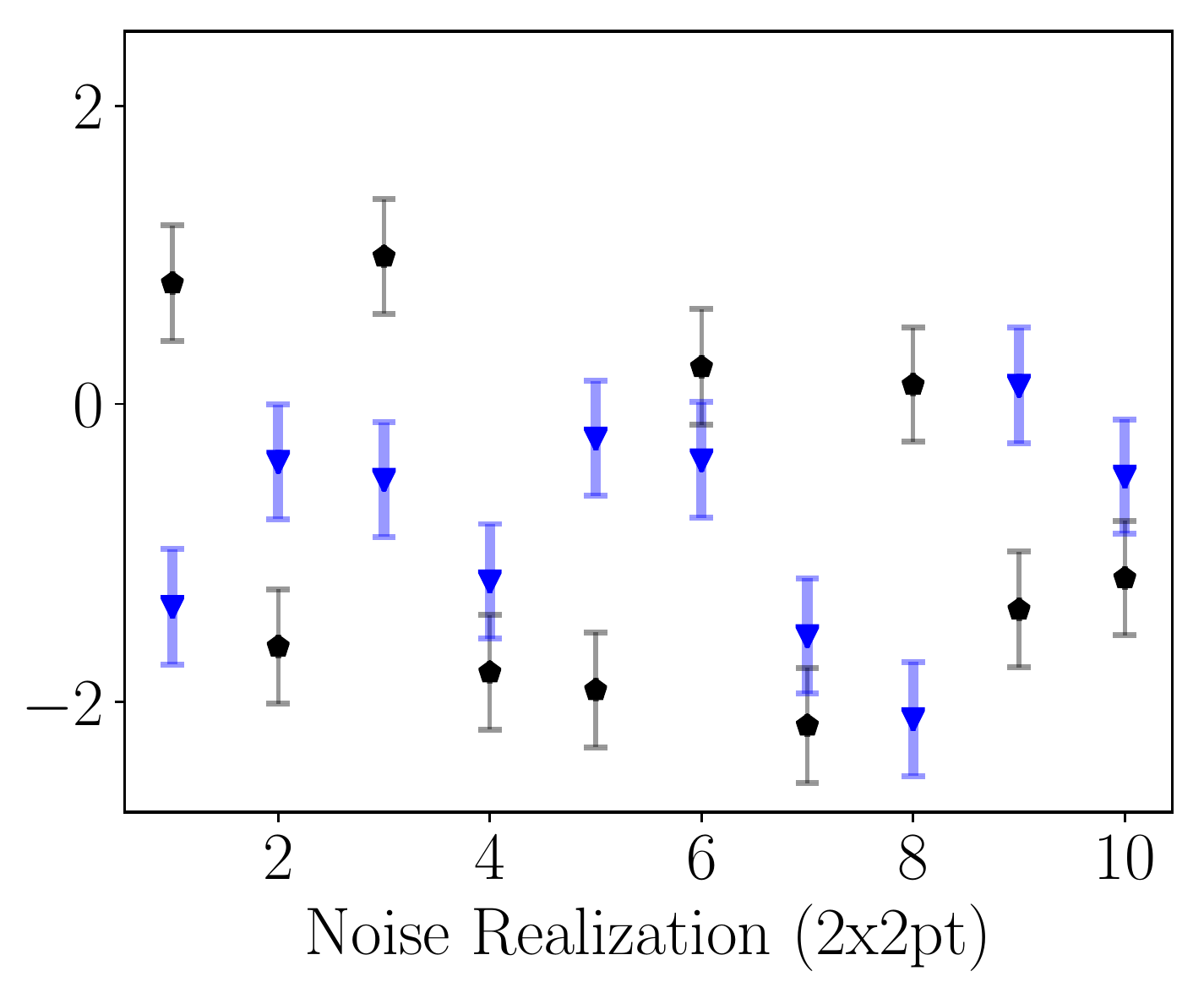}
\cprotect\caption{The panels present the comparison between the Bayesian evidence calculated using the Gaussian approximation and \verb+Polychord+'s results. Bias is defined as the difference for the natural logarithm of the Bayesian evidence. The left panel assumed the 3x2 pt data vector, while we restrict the analysis to galaxy-galaxy lensing and galaxy clustering in the right panel. The data vectors were randomly generated using a simulated DES-Y3 covariance. Triangle blue points with thick error bars show the results when the Gaussian approximation is made around the median of the chain, while black round points provide the results for the Gaussian estimation around the sample of chain with the best likelihood. The error bars reflect \verb+Polychord+'s claimed uncertainties.}
\label{fig:marco_calib_10dv}
\end{figure*}
\quad \quad  There is a significant difference in computational costs between running MCMC for parameter estimation and evaluating Bayesian evidence with nested sampling algorithms. The possibility of assessing evidence ratios using MCMC samples could, therefore, incentivize a more widespread use of such metric as well as make the recalibration of the Jeffreys scale a lot simpler. Such inference is, however, generically challenging in high-dimensional spaces (see ~\cite{Heavens:2017afc} and references within it). Recently,~\cite{Raveri:2018wln} proposed a Gaussian approximation to the posterior that can provide an estimate for the evidence ratio. For DES only chains, some partially constrained parameters are prior limited, which is an indication that the Gaussian approximation may fail. Nevertheless, we tested this approximation in few data vectors given the potential reward such a method could have brought to the ongoing DES-Y3 analysis and this work.

We have followed~\cite{Raveri:2018wln} closely, implementing the Gaussian approximation around either the best fit or the median of the MCMC chain. Initially, we have tested such a scheme in two noise realizations generated using an approximate DES-Y3 covariance (see table~\ref{tab:gaussian_test_models}). The use of DES-Y3 covariance matrix represents a best-case scenario given that more constraining data should make the Gaussian expansion to work better. For shear only, the approximation does not provide accurate Bayesian evidence ratios. Results were more encouraging for the 2x2pt and 3x2pt analyses, and we further examined such cases in eight additional noise realizations.  Results are shown in figure~\ref{fig:marco_calib_10dv}. Unfortunately, there are order unit biases that make the adoption of this approximation in our work unfeasible for even the most constraining 3x2pt analysis.
\FloatBarrier
\section{Halofit}
\label{section:appendixC}
\begin{figure}
\centering
\includegraphics[width=0.475\textwidth,height=0.485\textwidth]{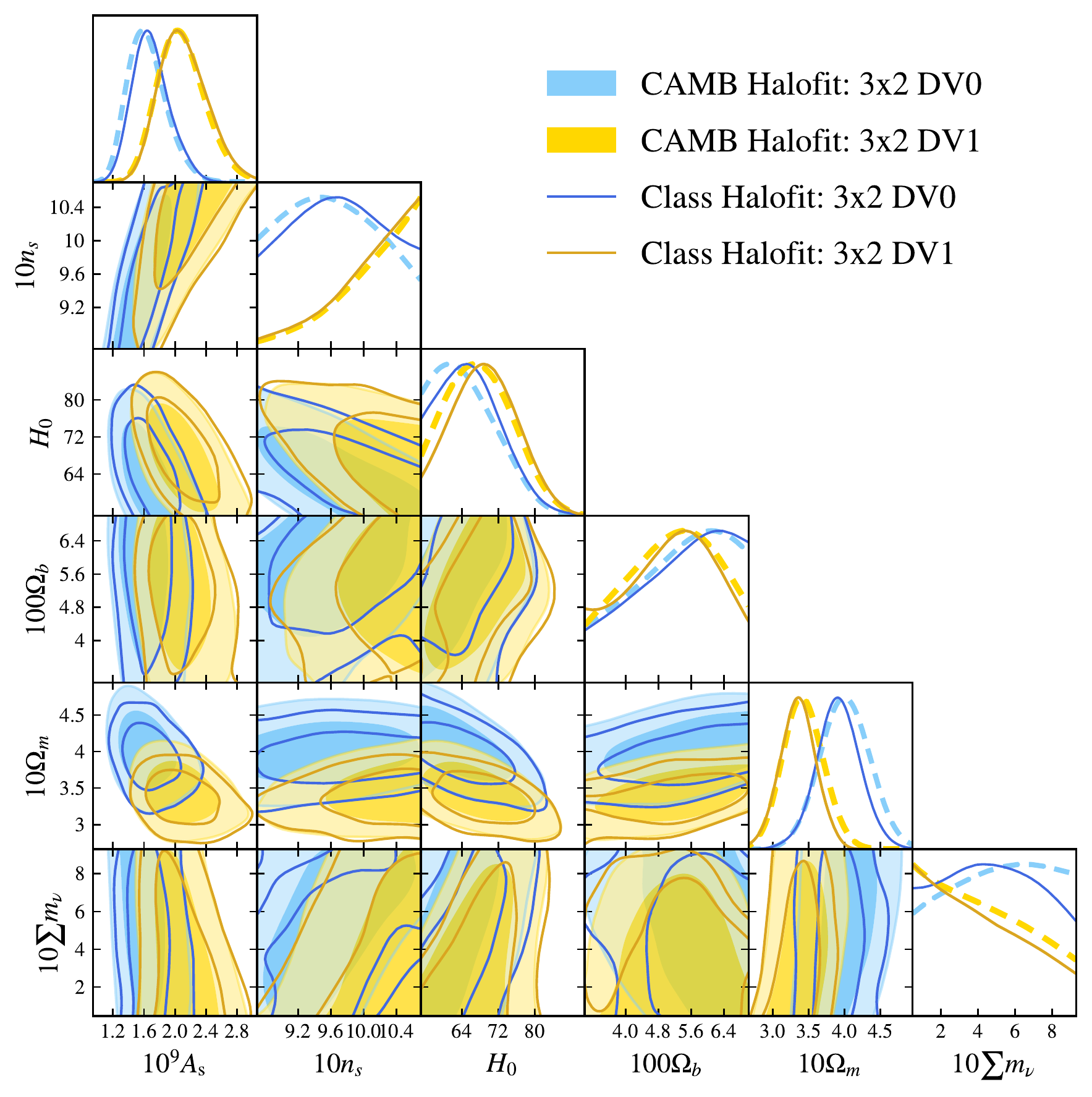}
\cprotect\caption{This figure compares the impact of the additional term that \verb+CLASS+ implements on the \verb+Halofit+ in comparison to the expression that \verb+CAMB+ assumes for the non-linear completion of the matter power spectrum. All MCMC chains adopted the Metropolis-Hasting sampler and \verb+CAMB+ code. Shades on the two-dimensional panels correspond to dashed lines on the one-dimensional posterior plots. The two 3x2pt data vectors - \verb'DV0' and \verb'DV1' were randomly generated around the default cosmology using a simulated DES-Y3 covariance. As expected, the posteriors differ the most on the volume of parameter space associated with high values for the sum of neutrino masses. Such discrepancy is also non-negligible on the one-dimensional $\Omega_m$ and $H_0$ marginalized posteriors.}
\label{fig:halofit_test}
\end{figure}
\quad \quad One practical issue has emerged in our sampler comparison that is related to implementation differences between \verb+CAMB+ and \verb+CLASS+ codes\footnote{\verb'CAMB' commit \verb+6884b632fa0bc2229a7bb18bf0b5d1f06c9913f2+ on the official \verb+GitHub+ repository \url{https://github.com/cmbant/CAMB}.  \verb+CLASS+ commit \verb+63f3cf18fad0061688b8bf95055765b4793f25c7+ on the official \verb+GitHub+ repository \url{https://github.com/lesgourg/class_public}}. The \verb+Cobaya+ pipeline version adopted in this work had only partial support to \verb+CLASS+, while \verb+CosmoLike+ is incompatible with \verb+CAMB+. Therefore, the Metropolis-Hasting and \verb+Polychord+ chains employed \verb+CAMB+ to evaluate the background comoving distances and the non-linear matter power spectrum, while \verb+Multinest+ and \verb+Emcee+ chains used \verb+CLASS+. We, consequently, tested the compatibility between these Boltzmann codes, and discrepancies in the \verb+Halofit+ formula were spotted.  

The original Takahashi \verb+Halofit+ formula for the non-linear matter power spectrum $\Delta^2(k) = k^3 P(k)/(2\pi^2)$ is given by
\begin{align}
\Delta^2(k) = \Delta^2_Q(k) + \Delta^2_H(k)\, .
\end{align}
The specific expression for $\Delta^2_Q(k)$ and $\Delta^2_H(k)$ can be found at~\citep{Takahashi:2012em}. Both \verb+Class+ and \verb+CAMB+ have updates to Takahashi formula that aims to provided better agreement against cosmology with massive neutrinos. We were unable to find the references in peer-reviewed journals for such updates. One of the new terms is, in \verb+Class+, the following 
\begin{align}\label{eqn:halofit_extra_term}
\Delta^2_Q(k) \to  \Delta^2_Q(k) \big\{1+f_\nu\big[0.977 -18.015 \times (\Omega_m - 0.3)\big]\big\} \,,
\end{align}
with $f_\nu \equiv \Omega_\nu/\Omega_m$. In \verb+CAMB+, on the other hand, the term proportional to $(\Omega_m - 0.3)$ does not exists; the impact of such factor is shown on figure~\ref{fig:halofit_test}.
\section{Nested Sampling}
\label{section:apxD}

\quad \quad Evaluation of the Bayesian evidence is possible with nested sampling algorithms~\citep{skilling2006}, and we will briefly review them in this appendix. Let $P(\vec{\theta} | \mathcal{H})$ be the prior distribution of the parameters $\vec{\theta}$ within a model $\mathcal{H}$, $\mathcal{L}$ be the likelihood distribution $P(\vec{d} | \vec{\theta}, \mathcal{H})$, and $\mathcal{E}$ be the evidence $P(\vec{d} | \mathcal{H})$. We define $X(\lambda)$ to be the fraction of the prior volume contained within the isolikelihood contour given by $P(\vec{d} | \vec{\theta}, \mathcal{H}) = \lambda$ as shown below
\begin{align}
X(\lambda) = \int_{\mathcal{L} > \lambda} d\vec{\theta} \, P(\vec{\theta} | \mathcal{H}) \, .
\end{align}

Nested sampling algorithms evaluate evidences via the one dimensional integral
\begin{align}
\mathcal{E} = \int_0^1 \mathcal{L}(X) dX \, .
\end{align}
This integration is performed by maintaining a set of live points, $n_\text{live}$, that samples a sequence of exponentially contracting volumes that respects that hard boundary $\mathcal{L} > \mathcal{L}_i$ at iteration $i+1$. The $\mathcal{L}_i$ value corresponds to the worse likelihood of all live points at iteration $i$, which is subsequently discarded and replaced by another point with $\mathcal{L} > \mathcal{L}_i$. Making this replacement efficient is the technically challenging part of the algorithm (see~\cite{Feroz:2013hea} and~\cite{2015MNRAS.453.4384H} for specific implementations). The set of discarded points are named dead points, and the discretization of the one dimensional evidence integral above is given by
\begin{align}
\mathcal{E} \approx \frac{1}{2}\sum_{i \in \, \text{dead}} \big(X_{i-1}-X_i\big) \times \mathcal{L}_i \,.
\end{align}
The precise $X_{i}$ volumes are unknown, but can be probabilistically estimated. To reconstruct the prior volume at the ith iteration, the algorithm sample $n_{\text{live}}$ times the uniform distribution spanning from 0 to $X_{i-1}$ and retrieve the maximum prior volume~\cite{skilling2006}. 

	The same procedure can also be used to calculate the KL divergence
\begin{align}
    \mathcal{D}_i \approx \frac{1}{2} \sum_{i \in \, \text{dead}} (X_{i-1}-X_{i}) \times \mathcal{L}_i \ln \bigg(\frac{\mathcal{L}_i}{\mathcal{E}}\bigg)  \, .
\end{align} 
This expression allows us to evaluate suspiciousness using the same nested sampling runs used to calculate evidence, and we have cross-check our numerical results for the KL divergence against the \verb+anesthetic+ package (run on the same chains)~\citep{Handley:2019mfs}. Finally, this section it also shows why the evaluation of the Surprise metric is challenging. The calculation of the relative entropy between datasets would require additional nested sampling runs where the ``prior'' would be one of the dataset's posteriors.  


\FloatBarrier
\bibliographystyle{mnras}
\bibliography{references.bib}
\end{document}